\documentclass{aa}

\usepackage{natbib}
\usepackage{graphicx}
\usepackage{xspace}
\usepackage{amsmath}
\usepackage{url}
\usepackage{hyperref}
\usepackage{inputenc}
\usepackage{color}

\def\mnras{MNRAS} 
\def\apj{ApJ} 
\def\aap{A\&A} 
\def\aj{AJ} 
\def\apjl{ApJ} 
\def\apjs{ApJS}

\def\nat{Nature} 
\def\prd{Phys.~Rev.~D}

\def\pasj{PASJ}
\def\jcap{JCAP}

\renewcommand{\d}{\mathrm{d}}

\newcommand\ch[1]{#1}
\newcommand\st[1]{}

\begin{document}

\title{Testing gravity with galaxy-galaxy lensing and redshift-space distortions using CFHT-Stripe 82, CFHTLenS and BOSS CMASS datasets} 
\titlerunning{Testing gravity from RSD and galaxy-galaxy lensing in CFHTLenS and CFHT-Stripe 82}
\authorrunning{E. Jullo et al.}

 \author{E. Jullo \inst{1}\thanks{\email{eric.jullo@lam.fr}} \and S. de la Torre \inst{1} \and M.-C. Cousinou \inst{2} \and S. Escoffier \inst{2} \and C. Giocoli \inst{3,4,5,6} \and R. Benton Metcalf \inst{4,5} \and J. Comparat \inst{7} \and H.-Y. Shan \inst{8} \and M. Makler \inst{9} \and J.-P. Kneib \inst{10,1} \and F. Prada \inst{11} \and G. Yepes \inst{12,13} \and S. Gottl\"ober \inst{14}}

 \institute{Aix-Marseille Univ, CNRS, CNES, LAM, Marseille, France 
              \and Aix Marseille Univ, CNRS/IN2P3, CPPM, Marseille, France
              \and Dipartimento di Fisica e Scienze della Terra, Universit\`a degli Studi di Ferrara, via Saragat 1, I-44122 Ferrara, Italy
              \and INAF - Osservatorio di Astrofisica e Scienza dello Spazio di Bologna, via Gobetti 93/3, I-40129 Bologna, Italy 
              \and Dipartimento di Fisica e Astronomia, Alma Mater Studiorum Universit\`a di Bologna, via Gobetti 93/2, I-40129 Bologna, Italy 
              \and INFN - Sezione di Bologna, viale Berti Pichat 6/2, I-40127 Bologna, Italy 
             \and Max-Planck-Institut f\"ur extraterrestrische Physik, Giessenbachstrasse 1, D-85748 Garching bei M\"unchen, Germany 
             \and Shanghai Astronomical Observatory (SHAO), Nandan Road 80, Shanghai 200030, China 
             \and Centro Brasileiro de Pesquisas F\'isicas, Rio de Janeiro, RJ 22290-180, Brasil 
             \and Institute of Physics, Laboratory of Astrophysics, Ecole Polytechnique F\'ed\'erale de Lausanne (EPFL), Observatoire de Sauverny, 1290 Versoix, Switzerland
              \and Instituto de Astrof\'isica de Andaluc\'ia (CSIC), Glorieta de la Astronom\'ia, E-18080 Granada, Spain 
              \and Departamento de F\'isica Te\'orica, M\'odulo 15, Universidad Aut\'onoma de Madrid, 28049 Madrid, Spain 
             \and Centro de Investigaci\'on Avanzada en F\'isica Fundamental (CIAFF), Universidad Aut\'onoma de Madrid, 28049 Madrid, Spain 
             \and Leibniz-Institut f\"ur Astrophysik (AIP), An der Sternwarte 16, D-14482 Potsdam, Germany 
}
             
\maketitle             
\begin{abstract}

The combination of Galaxy-Galaxy Lensing (GGL) and Redshift Space Distortion of galaxy clustering (RSD) is a privileged technique to test General Relativity predictions, and break degeneracies between the growth rate of structure parameter $f$ and the amplitude of the linear power-spectrum $\sigma_8$. We perform a joint GGL and RSD analysis on 250 sq. degrees using shape catalogues from CFHTLenS and CFHT-Stripe 82, and spectroscopic redshifts from the BOSS CMASS sample. We adjust a model that includes non-linear biasing, RSD and Alcock-Paczynski effects. We find $f(z=0.57) =0.95\pm0.23$, $\sigma_8(z=0.57)=0.55\pm0.07$ and $\Omega_{\rm m} = 0.31\pm0.08$, in agreement with Planck cosmological results 2018. We also estimate the probe of gravity $E_{\rm G} = 0.43\pm0.10$ in agreement with $\Lambda$CDM-GR predictions of $E_{\rm G} = 0.40$. This analysis reveals that RSD efficiently decreases the GGL uncertainty on $\Omega_{\rm m}$ by a factor of 4, and by 30\% on $\sigma_8$. We use an N-body simulation supplemented by an abundance matching prescription for CMASS to build a set of overlapping lensing and clustering mocks. Together with additional spectroscopic data, this helps us to quantify and correct several systematic errors, such as photometric redshifts. We make our mock catalogues available on the Skies and Universe database\footnote{\url{http://www.skiesanduniverses.org}}. 

\end{abstract}

\section{Introduction}

Since its inception, General Relativity theory (GR) has been constantly tested, starting with observations in the Solar system and in our Galaxy \citep[see e.g.][]{damour2000}. 
Today at cosmological scales, the advent of wide field survey experiments yields very high precision measurements in both the early and late ages of the universe. A Universe dominated by Cold Dark Matter and a cosmological constant in the context of general relativity (hereafter $\Lambda$CDM-GR model) reproduces all these observations with very high accuracy and for this reason, the model is often referred to as the standard reference model. 

However, some slight tensions are  emerging between predictions based on the Cosmic Microwave Background measurements from the Planck mission at redshift $z=1089$ and measurements at redshifts $z<1$ obtained from galaxy clustering or gravitational lensing. In particular with Planck, the amplitude of the matter power spectrum $\sigma_8$ is larger and the Hubble constant $H_0$ is smaller than what is estimated at redshifts $z<1$ at about $2\sigma$ Confidence Level (hereafter C.L.) \citep{planck2016,beutler2014,alam2016,hildebrandt2017,des2017}. \ch{Although systematic errors in the analyses can explain a significant fraction of these discrepancies, they might nonetheless suggest some issue with our understanding and modeling of the universe's expansion, or of the large-scale structure formation probed by galaxy clustering and gravitational lensing}. \st{This opens the tantalizing possibility of replacing $\Lambda$CDM-GR and its cumbersome dark sector by an alternative model of gravity.}

The common approach to test  $\Lambda$CDM-GR at cosmological scales is either to measure the universe's expansion history $H(z)$ \citep[e.g.][]{betoule2014,alam2016,magana2015}, or to measure the growth of structures traced by the velocity or density fields in redshift space \citep[e.g.][]{delatorre2013,tully2016,martinet2018}. In this paper, we combine GGL and RSD to test \ch{both aspects simultaneously} \st{the consistency between the velocity and density fields} at redshift $z=0.57$. \st{At large scales, t} \ch{T}he amplitude of GGL measurements \st{probes the density field} \ch{is sensitive to $H(z)$ and the density field}, whereas RSD probes the growth of structure $f(z)$ through galaxy peculiar velocities. The combination of these 2 observables has demonstrated its effectiveness at isolating the independent effects of the growth rate of structure $f(z)$, \st{and} the amplitude of the matter power spectrum $\sigma_8$, \ch{and the dark energy equation of state parameter $w$ involved in $H(z)$ calculation} \citep{simpson2013,delatorre2017,joudaki2017}.

\cite{zhang2007} proposed an alternative method to test deviations to GR. Assuming small scalar perturbations around the FLRW metric in the conformal Newtonian gauge $ds^2 = -a(\tau)^2[1+2\Psi]d\tau^2+ a(\tau)^2[1-2\Phi]d{\bf x}^2$, where $a$ is a scale factor, $\tau$ is the conformal time, and $\bf x$ are comoving coordinates, they proposed a statistics $E_{\rm G}$ sensitive to the gravitational slip between the 2 gravitational potentials $\Phi$ and $\Psi$

\begin{equation}
\langle E_{\rm G}  \rangle = \left[ \frac{\nabla^2 (\Psi - \Phi)}{3H_0^2 a^{-1} f \delta} \right] \,.
\end{equation}

\noindent where \st{a is the scale factor at the redshift of the lens}\ch{all quantities are estimated at the redshift of interest. \citet{reyes2010} proposed an associated observational estimator $E_{\rm G} = \Upsilon_{\rm gm}/\beta\Upsilon_{\rm gg}$ (see below in section \ref{sec:eg}), which converges to $\langle E_{\rm G} \rangle$ in the large scale limit where the galaxy bias $b$ and the distortion parameter $\beta = f/b$ converge to constant values. The small scale filtered galaxy-matter cross-correlation $\Upsilon_{\rm gm}$ probed with GGL is sensitive to both $b$ and $\nabla^2(\Phi - \Psi)$  since photons traverse equal quantity of space and time. The galaxy-velocity cross-correlation $\beta \Upsilon_{\rm gg}$ probed with RSD is sensitive to galaxy bias and the Newtonian potential $\Psi$. In GR and in absence of anisotropic stress, $\Phi = -\Psi$ so lensing is sensitive of $2 \nabla^2 \Phi$.  In the linear regime, the Poisson equation relates the potential to the matter density contrast $\delta$, such that $\nabla^2 \Phi = -\nabla^2 \Psi =\frac{3}{2} \Omega_{m0} H_0^2 a^{-1} \delta$.} This estimator therefore converges to  $E_{\rm G} = \frac{\Omega_{m0}}{f}$ in the standard model. 

\ch{In their seminal paper, \cite{zhang2007} predicted deviations from GR with 4 alternative models. Apart from the scalar-tensor models, which introduce a wavelength-dependent difference between dynamical and lensing power-spectra, all other models just add at most 10\% deviations compared to GR predictions. \cite{leonard2015} also explored other models and reached similar conclusions. Most importantly, they found that details of the analysis (e.g. integration length along the line of sight for projected estimators), could mimic similar deviations, thus the need for a careful study of these biases. In any case with 20\% to 30\% precision, current datasets are not yet at the level of accuracy required to observe these deviations, and unsurprisingly no deviation to GR predictions has been detected so far \citep{blake2016,pullen2016,delatorre2017,alam2016,amon2017}.}

Nowadays, cosmological analyses require measurements with exquisite control of systematic errors, at all levels from data acquisition to cosmological model inference. The wide range of expertise needed to reach the requirements is demonstrated by the size of the on-going and forthcoming cosmological experiments such as Dark Energy Survey \citep{des2005}, the Kilo Degree Survey \citep{hildebrandt2017}, the Hyper-Suprime Cam survey \citep{aihara2018}, the extended Baryonic Oscillation Sky Survey \citep{dawson2013}, the Prime Focus Spectrograph project \citep{sugai2012}, the Dark Energy Survey Instrument project \citep{desi2016a,desi2016b}, the Large Scale Synoptic Telescope \citep{lsst2012} and the Euclid mission \cite{laureijs2011}, 

In this paper, we extend the \cite{leauthaud2017} analysis (hereafter L17), by adding RSD measurements of CMASS galaxies from the Baryon acoustic Oscillation Spectroscopic Survey (BOSS) to GGL measurements in the CFHT-Stripe 82 and CFHT-LS fields. Thanks to refined simulations, we precisely quantify systematic errors, and thus manage to reconcile \ch{real and simulated measurements of clustering and lensing} \st{measured in real data and in simulations}. The work presented here builds on the theoretical model and joint RSD and GGL analysis developed in \cite{delatorre2017} (hereafter DLT17).

The outline of the paper is as follows. First we present our galaxy bias model, and its inclusion in standard clustering and lensing estimators. Next, we present our datasets and measurement estimators. Our tests on simulations are presented in Section \ref{sec:simus}, and our estimates of the cosmological parameters in Section \ref{sec:results}. Finally, we present our measurement of $E_{\rm G}$ and conclude. Systematic errors are discussed in the Appendix. Unless otherwise mentioned, we express the GGL projected densities $\Sigma_{\rm gm}$ and distances in comoving coordinates. We assume the fiducial  $\Lambda$CDM-GR cosmology with flat universe, $\Omega_{\rm m} = 0.31$, $h=0.6777$, $\Omega_{\rm b} = 0.048$, $\sigma_8=0.82$ \citep{planck2016}.

\section{Method}

In the following, we compute the RSD two-point galaxy correlation functions in configuration space. We decompose the three-dimensional galaxy separation vector ${\bf s}$ into polar \ch{$(s,\mu)$ or cartesian $(r_{\rm p},\pi)$} coordinates in the frame defined by the line-of-sight and the normal to it, where $s$ is the norm of \st{the separation-vector} ${\bf s}$, $\mu$ is the cosine of the angle between ${\bf s}$ and the line-of-light  \st{the separation-vector direction}, $\pi$ and $r_{\rm p}$ are the \ch{projections of ${\bf s}$ on the line-of-sight and its normal respectively.} \st{and $\pi$ is the distance parallel to the line-of-sight}. \ch{In the flat-sky approximation, the transformation between cartesian and polar coordinates is $\mu = \pi / s$, $r_{\rm p} = \sqrt{s^2 - \pi^2}$  \cite{fisher1994}}. Conversely, the GGL formalism is defined in real space, where the separation vector $\bf r$ is decomposed into cartesian coordinates $(R,\chi)$\ch{, where $\chi$ and $R$ are the projections of ${\bf r}$ on the line-of-sight and its normal respectively.} In GGL, the radial window function of integration is hundreds of $h^{-1}\ {\rm Mpc}$, and the effects of RSD can safely be neglected \citep{baldauf2010}.  Hereafter, we will assume that $R$ in the model corresponds to $r_{\rm p}$ in the observations.

\subsection{Galaxy bias model}
In this work, we want to measure the growth rate $f$ and the amplitude of the matter power spectrum $\sigma_8$ with galaxy-galaxy lensing and galaxy-clustering measurements. These measurements are not typically estimated at the same scale. While the GGL signal is typically studied in the range of transverse distances $0.1<r_{\rm p}<20 h^{-1}\ {\rm Mpc}$, RSD studies focus on the range $10<r_{\rm p}<100 h^{-1}\ {\rm Mpc}$. This difference is due to observational and modeling considerations. In order to maximize the overlap between these two observables in the non-linear regime, we adopt the 4$^{th}$-order perturbation model in the initial density field as proposed by \cite{mcdonald2009}. Assuming homogeneity and isotropy in the density field \st{as well as symmetry and equivalence in the calculations}, \ch{they derived the following expression for the halo-matter power-spectrum 
\begin{multline}
P_{\rm gm}(k) = b_1 P_{\delta \delta}(k) + b_2 P_{b_2,\delta}(k) + \\ b_{s^2} P_{b_{s^2},\delta}(k) + b_{\rm 3nl} \sigma_3^2(k) P_{\rm lin}(k)\,,
\end{multline}
\noindent where $P_{\delta \delta}$ and $P_{\rm lin}$ represent the non-linear and linear matter power-spectra respectively,  $P_{b_2,\delta}$ and $P_{b_{s^2},\delta}$ are the one-loop power-spectra between the density field $\delta$, its derivative and the variance of the tidal tensor field $s({\bf x})$. The term $b_{\rm 3 nl} \sigma_3^2(k)$ includes various 3$^{rd}$-order terms of the galaxy bias model  \citep[see][for more details]{mcdonald2009}. Assuming coevolution between the halo and the matter density fields, as well as the bias being purely local in Lagrangian space at initial conditions, \cite{baldauf2012} computed the 2$^{nd}$-order halo density field in both Eulerian and Lagrangian space and found the relation $b_{s^2} = -4/7(b_1 -1)$.  Under the same assumptions as above to compute $b_{s^2}$, \cite{saito2014} obtained the relation $b_{\rm 3nl} = 32/315(b_1-1)$. The analytical expressions for all these terms are given in Appendix A of DLT17. 
}

\subsection{Galaxy-galaxy lensing model}

The measured GGL differential excess surface density is defined as
\begin{equation}
\Delta \Sigma_{\rm gm} (R) = \overline{\Sigma}_{\rm gm}(R) - \Sigma_{\rm gm}(R)
\end{equation}
\noindent where  the mean  projected surface density can be read as

\begin{equation}
\overline{\Sigma}_{\rm gm}(R) = \frac{2}{R^2} \int_{0}^{R} \Sigma_{\rm gm}(r)\, r\, \d r
\end{equation}
and $\Sigma_{\rm gm}(R)$ is the projected surface density defined as a function of the galaxy-matter cross-correlation function \citep{guzik2001,johnston2007}
\begin{equation}
\Sigma_{\rm gm}(R) = \Omega_{\rm m} \rho_c \int_{-\infty}^{\infty} \left( 1+ \xi_{\rm gm}\left( \sqrt{R^2 + \chi^2} \right) \right)\, \d \chi
\end{equation}
\noindent where the mean matter density $\rho_{\rm m} = \Omega_{\rm m} \rho_c = 3 \Omega_{m0} H_0^2 / 8\pi {\rm G}$ is constant in comoving coordinates. The galaxy-matter cross-correlation function $\xi_{\rm gm}$ is obtained from the Fourier Transform of the galaxy-matter power-spectrum $P_{\rm gm}(k)$ defined above.

In practice, we use an \textsc{FFTLog} unbiased Hankel transform with parameter $\mu = \frac{1}{2}$ in logarithmic space to perform the Fourier Transform\footnote{http://casa.colorado.edu/ajsh}. We truncate the power-spectrum at $k_{\rm min}=10^{-5}$ and $k_{max} = 1000$  in order to minimize cut-off aliasing during the FFT operation, and we spline-interpolate the resulting correlation function to obtain the desired binning.

\subsection{Redshift space distortions model}

In this work, we use the \cite{taruya2010} model  to describe the RSD effect. In the ideal case where galaxies are perfect tracers of the matter density field, this model takes the form:
\begin{multline}
P^s(k,\mu)= D(k\mu\sigma_v)\big[ P_{\delta\delta}(k) +2\mu^2f P_{\delta\theta}(k) + \mu^4f^2P_{\theta\theta}(k)+ \\ C_A(k,\mu,f)+C_B(k,\mu,f) \big]\,,
\end{multline}
\noindent where $\theta$ is the divergence of the velocity field defined as $\theta = -\nabla {\bf \cdot v}/(aHf)$. $P_{\delta\delta}$,  $P_{\theta\theta}$ and $P_{\delta\theta}$  are respectively the non-linear matter density, velocity divergence, and density-velocity divergence  power-spectra; $C_A(k,\mu,f)$ and $C_B(k,\mu,f)$ terms derive from the general anisotropic power-spectrum of matter and their expressions are given in \cite{taruya2010} and \cite{delatorre2012}. 

The damping function $D(k\mu\sigma_v)$, essentially (but not only) \ch{describes the Fingers-of-God effect on the two-point correlation function}, and we model it as a Lorentzian \ch{damping} in Fourier space, i.e.
\begin{equation}
D(k,\mu,\sigma_v) = (1+k^2\mu^2\sigma_v^2)^{-1}\,,
\end{equation}
\noindent where $\sigma_v$ represents an effective pairwise velocity dispersion that we fit for and then treat as a nuisance parameter. 

This model can be generalized to the case of biased tracers, by including our bias model. Hence, we obtain \citep{beutler2014,gilmarin2014}
\begin{multline}
P^s_{\rm g}(k,\mu) = D(k\mu\sigma_v) \big[ P_{\rm gg}(k) + 2\mu^2fP_{\rm{g} \theta} + \mu^4f^2 P_{\theta\theta}(k) + \\
C_A(k,\mu,f,b_1) + C_B(k,\mu,f,b_1) \big]
\label{eq:psg}
\end{multline}
\noindent where,
\begin{multline}
P_{\rm gg}(k) = b_1^2 P_{\delta\delta}(k)+2b_2b_1P_{b_2,\delta}(k)  +2b_{s^2}b_1P_{b_{s^2},\delta}(k) \\ +b_2^2 P_{b_2,b_2}(k) + 2b_2b_{s^2}P_{b_2,b_{s^2}}(k)+ b_{s^2}^2 P_{b_{s^2},b_{s^2}}(k) \\ + 2b_1b_{\rm 3nl}\sigma_3^2(k)P_{\rm lin}(k)
\end{multline}
\begin{multline}
P_{\rm{g}\theta}(k)=b_1P_{\delta\theta}(k)+b_2P_{b_2,\theta}(k)+\\b_{s^2}P_{b_{s^2},\theta}(k)+b_{\rm 3nl}\sigma_3^2(k)P_{\rm lin}(k)\,.
\end{multline}
In the above equations $P_{b_2,\delta}$, $P_{b_{s^2},\delta}$, $P_{b_2,b_2}$, $P_{b_2,b_{s^2}}$, $P_{b_{s^2},b_{s^2}}$ and $\sigma_3^2(k)$ are 1-loop integrals, of which analytical expressions can be found in Appendix A of DLT17. We compute the linear matter power-spectrum $P_{\rm lin}$ using the \textsc{class} Bolzmann code \citep{lesgourgues2011}, and the non-linear matter power-spectrum $P_{\delta\delta}$ using the semi-analytic prescriptions HALOFIT \citep{smith2003,takahashi2012}. To predict the velocity spectra $P_{\theta\theta}$ and $P_{\delta\theta}$, we use the nearly universal fitting functions from \cite{bel2018}, already used in DLT17 and \cite{pezzotta2017}. They are built such that they converge to $P_{\rm lin}$ at large scales, but reproduce non-linearities at small scales. \cite{pezzotta2017} highlighted that adding a redshift dependency with $\sigma_8(z)$ such that
\begin{equation}
P_{\theta\theta}(k) = P_{\rm lin}(k)\exp[-k\,p_1\,\sigma_8^{p_2}(z)]
\end{equation}
and 
\begin{equation}
P_{\delta\theta}(k) = \sqrt{P_{\delta\delta}P_{\rm lin}(k)\exp[-k\,p_3\,\sigma_8^{p_4}(z)]}\,,
\end{equation}
\noindent was helping. The coefficients $(p_1=1.906,p_2=2.163,p_3=2.972,p_4=2034)$ were deduced from a fit to measurements performed on the DEMNUni simulations (Dark Energy and Massive Neutrino Universe). These 2 fitting functions are accurate within 5\% to the measurements in simulations, and appear to be insensitive to the presence of neutrinos \citep{carbone2016}. The overall degree of non-linearity in these terms is therefore solely controlled by $\sigma_8(z)$, which is left free when fitting the model to observations. \ch{Although these fitting functions possibly duplicate a fraction of the high-order modes included in the perturbation theory model above, we demonstrate in DLT17 and in Section~\ref{sec:simus} below that it does not bias significantly our cosmological estimates given data uncertainties.}

Finally, we obtain the multipole moments of the anisotropic correlation functions in configuration space
\begin{equation}
\xi_\ell^s(s) = i^\ell \int \frac{k^2}{2\pi^2} P_\ell^s(k)j_\ell(ks) \d k
\end{equation}
\noindent where $j_\ell(x)$ is the spherical Bessel function and  $P_\ell^s(k)$ is the anisotropic power-spectrum multipole moment of order $\ell$ defined as 
\begin{equation}
P_\ell^s(k) = \frac{2\ell + 1}{2}\int_{-1}^1 P_{\rm g}^s(k,\mu)L_\ell(\mu) \d \mu\,.
\end{equation}
\noindent where $L_\ell(x)$ are the Legendre polynomial of order $\ell$. 

\st{Generally in RSD analysis,} \ch{At linear scales,} $f$ and $\sigma_8$ are degenerate, but \ch{extending to non-linear scales with} the \cite{taruya2010} model, $b_1^2f\sigma_8^4$, $b_1f^2\sigma_8^4$ and $f^3\sigma_8^4$ appear in the calculation of the correction terms $C_A$ and $C_B$, and hence help break the degeneracy. Accordingly, in our model ($f$, $b_1$, $b_2$, $\sigma_v$, $\sigma_8$) are treated as separate parameters in the fit \citep{delatorre2012}.\st{, and we provide marginalized constraints on the derived parameters, such as $f\sigma_8$.} 

\subsection{Spectroscopic redshift uncertainties}

It is worth mentioning that redshift errors can potentially affect the anisotropic RSD signal. To the extent that they are Gaussian distributed, they have the same effect as galaxy random motions in virialised objects. We model them by multiplying the anisotropic power-spectrum by the Fourier transform of a Gaussian damping function of the form 
\begin{equation}
G(k,\mu,\sigma_z) = \exp \left( -\frac{k^2\mu^2\sigma_z^2}{2} \right)\,,
\end{equation}
\noindent such that our predicted signal can be finally written as:
\begin{equation}
\widehat{P_{\rm g}^s} = G(k,\mu,\sigma_z) P_{\rm g}^s\,. 
\end{equation}

\cite{bolton2012} measured the error on the estimated spectroscopic velocities, thanks to multiple observations of the same CMASS galaxies, and found approximately $\delta_v = 32\ {\rm km\ }{\rm s}^{-1}$, which translates to $\sigma_z = 0.797\ h^{-1}\ {\rm Mpc}$ in comoving distances at redshift $z=0.57$ with our fiducial cosmology. This effect is therefore negligible, but we included it, in order to have a cleaner estimate of $\sigma_v$.

\subsection{Suppressing small scale modeling uncertainties}
\label{sec:upsilon}

Although considered as sufficient for galaxy-clustering analysis, we find that our weak-lensing model deviates from our measurements with simulated data at scales $r_{\rm p} \sim 3\ h^{-1}\ {\rm Mpc}$ (see Fig. 7 in DLT17). In order to damp the contribution of any signal below a given cut-off radius $R_0$, we compute the annular differential excess surface density (ASAD) estimator from the data \citep{baldauf2010}. For the lensing observable $\Delta \Sigma_{\rm gm}(r_{\rm p})$, it is given by

\begin{equation}
\Upsilon_{\rm gm}(r_{\rm p}, R_0) = \Delta \Sigma_{\rm gm}(r_{\rm p}) - \left( \frac{R_0}{r_{\rm p}} \right)^2 \Delta \Sigma_{\rm gm} (R_0)\,,
\end{equation}

\noindent and for the galaxy-clustering 

\begin{multline}
\Upsilon_{\rm gg}(r_{\rm p},R_0) = \rho_c \left[ \frac{2}{r_{\rm p}^2} \int_{R_0}^{r_{\rm p}} \d r\ r\ w_{\rm p}(r) - \right. \\ 
\left. w_{\rm p}(r_{\rm p}) + \frac{R_0^2}{r_{\rm p}^2}  w_{\rm p}(R_0) \right].
\end{multline}

\noindent These two estimators will be useful to estimate $E_{\rm G}$ in the following. \ch{We derive the projected correlation $w_{\rm p}(r_{\rm p})$ from the projection of the multipole decomposition of the correlation function in redshift space $\xi_\ell^s(s)$

\begin{multline}
w_{\rm p}(r_{\rm p}) = 2 \sum_{\ell=0}^2 \alpha_{2\ell} \int_0^{\pi_{max}} \d \pi\ \xi_{2\ell}^s \left( \sqrt{r_{\rm p}^2+\pi^2}\right) \\ \times L_{2\ell} \left(\frac{\pi}{\sqrt{r_{\rm p}^2+\pi^2}}\right).
\end{multline}

\noindent   The $\alpha_{2\ell}$ coefficients are given in  \citet{baldauf2010} :

\begin{eqnarray}
\alpha_0(\beta) &=& 1+ \frac{2}{3}\beta + \frac{1}{5}\beta^2 \,; \\
\alpha_2(\beta) &=& \frac{4}{3}\beta + \frac{4}{7}\beta^2 \,;\\
\alpha_4(\beta) &=& \frac{8}{35}\beta^2\,.
\end{eqnarray}

\noindent We integrate \ch{along the line-of-sight} up to $\pi_{max} = 40 h^{-1}\ {\rm Mpc}$, to match the integration length used with the data (see the estimators section~\ref{sec:estimators}). According to \cite{singh2018}, they found consistent results, whether they use $\pi_{max} = 50 h^{-1}\ {\rm Mpc}$ or $100 h^{-1}\ {\rm Mpc}$. Given the low number CMASS galaxies in this analysis, we set $\pi_{max} = 40 h^{-1}\ {\rm Mpc}$ to minimize the noise.}

The ASAD can also be predicted from theory. For the lensing part, $\Upsilon_{\rm gm}(r_{\rm p},R_0)$ is obtained by filtering the cross-correlation function $\xi_{\rm gm}(r)$ 
\begin{equation}
\Upsilon_{\rm gm} (r_{\rm p}, R_0) = \int_0^\infty \xi_{\rm gm} (r) W_\Upsilon(r,r_{\rm p}, R_0) \d r\,,
\label{eq:upssgm}
\end{equation}
\noindent with the window function  $W_\Upsilon(x,r_{\rm p},R_0)$ \citep{baldauf2010} defined as:
\begin{multline}
\tiny
W_\Upsilon(x,r_{\rm p},R_0) = \\ \quad \frac{4x}{r_{\rm p}^2}\left( \sqrt{x^2-R_0^2}\,\Theta(x-R_0) -   \sqrt{x^2-r_{\rm p}^2}\,\Theta(x-r_{\rm p})\right)   \\
 - \frac{2x}{r_{\rm p}^2}\left(\frac{r_{\rm p}^2\,\Theta(x-r_{\rm p})}{\sqrt{x^2-r_{\rm p}^2}}-\frac{R_0^2\,\Theta(x-R_0)}{\sqrt{x^2-R_0^2}}\right)\,,
\end{multline}
\noindent where $\Theta(x)$ is the Heaviside step function.  In a similar manner, we compute  $\Upsilon_{\rm gg}(r_{\rm p},R_0)$ by simply replacing $\xi_{\rm gm}(r)$ by $\xi_{\rm gg}(r)$ in equation~\ref{eq:upssgm}. Note that we include RSD effect in the calculation of $\Upsilon_{\rm gg}(r_{\rm p})$. In both cases, we integrate in logarithmic scale up to $r_{max} = 100 h^{-1}\ {\rm Mpc}$. 

Note that we do not include intrinsic alignment in our modeling. This choice is motivated by the marginal constraints obtained in \cite{joudaki2017} on the amplitude of this effect $A_{IA} = 1.67^{+0.50}_{-0.49}$, with small scale cut on $\gamma_t$ at $\theta > 12$ arcmin. Since we apply the small scale $\Upsilon$ filter, we anticipate very little constraint on this parameter as well, at a significant additional computing cost. 

\subsection{Alcock-Paczynski effect}

We need to mention that additional distortions can occur in the correlation functions, due to possible differences between the true and the fiducial cosmological models used to compute the distances. This effect was first identified by \citep[][AP]{alcock1979} as a means to constrain the cosmological model. However these distortions are degenerate with the RSD effect and considerably limit the constraining power of the AP effect \citep{ballinger1996,matsubara1996}. Fortunately, the scale-dependence of the AP and RSD effects differ, and thus allow us to break this degeneracy \citep{seo2003,blake2011,chuang2012}. 

In this work, we adopt the AP model proposed by \cite{xu2013}. The isotropic and anisotropic distortions are expressed respectively as

\begin{equation}
\alpha = \left( \frac{D_A^2}{{D'_A}^2} \frac{H'}{H} \right)^{1/3}
\end{equation}

\begin{equation}
1+\epsilon = \left( \frac{D'_A}{D_A} \frac{H'}{H} \right)^{1/3}
\end{equation}

\noindent where quantities computed with the fiducial cosmology as denoted with primes. Those parameters modify the transverse and the radial distances such that 

\begin{equation}
\pi' = \alpha (1+\epsilon)^2 \pi
\end{equation}

\begin{equation}
r_{\rm p}' = \alpha (1+\epsilon)^{-1} r_{\rm p}\,.
\end{equation}

Given these distortions, the observed redshift-space monopole and quadrupole expressed in configuration space become

\begin{equation}
\xi_0'(s') = \xi_0(\alpha s) + \frac{2}{5} \epsilon \left[ 3\xi_2(\alpha s) + \frac{\d \xi_2(\alpha s)}{\d \ln(s)} \right] 
\end{equation}

\begin{multline}
\xi_2'(s') = 2 \epsilon \frac{\d \xi_0(\alpha s)}{\d \ln(s)} + \left(1+\frac{6}{7}\epsilon \right) \xi_2(\alpha s) + \frac{4}{7} \epsilon \frac{\d \ln(\alpha s)}{\d \ln(s)} \\
 + \frac{4}{7} \epsilon \left[ 5 \xi_4(\alpha s) + \frac{\d \xi_4(\alpha s)}{\d \ln(s)} \right]\,,
\end{multline}

\noindent The GGL estimator becomes 

\begin{equation}
\Upsilon'_{\rm gm} (R') = \Upsilon_{\rm gm}[\alpha(1+\epsilon)^{-1} R]\,.
\end{equation}

\section{Data}

\ch{In our GGL analysis, the lenses are the CMASS galaxies, and the sources are galaxies in the CFHTLens and CFHT-Stripe 82 weak-lensing catalogues. Lenses have spectroscopic redshifts, and sources have photometric redshifts. For each lens, we can then discard all uncorrelated foreground sources, and use the background sources to estimate the lensing signal. The final GGL measurement is the average of the signals for each lens. }

\subsection{Weak Lensing Datasets}

\subsubsection{The CFHTLens catalogue}

In 2013, the CFHTLenS team released a public weak lensing catalogue covering an area of 154 sq. degrees in 4 wide fields  (W1, W2, W3 \& W4) \citep{erben2013,heymans2012}. So far, the depth of the input CFHT Legacy Survey imaging is unrivaled, with a $5\sigma$ point source limiting magnitude $i_{\rm AB} \sim 25.5$. The \textsc{lensfit} algorithm is used to measure the shape of every object detected with $i_{\rm AB} < 24.7$. Then, we select galaxies with good shape measurement ($\textsc{fitclass}=0$ and $\textsc{weight}>3$). 

Photometric redshifts are obtained from five optical band photometry {\it u, g, r, i, z} and reach a precision of about 5\% up to $z \sim 1$ \citep{hildebrandt2012}. GGL measurements can be significantly biased by inaccurate photometric redshifts \citep{nakajima2012}. We compute the photometric redshift bias estimator $\langle b_z \rangle$, based on spectroscopic and photometric catalogues matched in position, and averaged over the CMASS redshift distribution (see Appendix details). \ch{Since the spectroscopic calibration sample is significantly shallower than the photometric sample, we have to} discard galaxies fainter than the 90\% completeness limit of the spectroscopic sample (see below), i.e. we only keep galaxies brighter than $i_{\rm AB} < 24$. After this selection, we obtain  $\langle b_z \rangle =+0.003\pm0.003$, $\langle b_z \rangle = -0.014\pm0.004$ and $\langle b_z \rangle =+0.022\pm0.003$ in fields W1, W3 and W4 respectively. We discard field W2 because it only contains 200 CMASS galaxies on its Northern edge.

Our final catalogue contains 3.5 millions galaxies over an effective area of about 127 sq. degrees. The galaxy density\footnote{We use the definition $n_{\rm eff} = \frac{1}{\Omega} \frac{(\sum w_i)^2}{\sum w_i^2}$ \citep{heymans2012}, where $w_i$ is a galaxy weight, and $\Omega$ is the opening angle} is $n_{\rm eff} = 7.0$ galaxies arcmin$^{-2}$. The median redshift is $z_{med} = 0.70$.

\subsubsection{The CFHT-Stripe 82 catalogue}
\label{sec:cs82cat}

The CFHT-Stripe 82 survey \citep[CS82,][]{moraes2014} is an $i$-band imaging survey containing 173 tiles. It covers about 160 sq. degrees of the SDSS Stripe 82 region, with a 5-$\sigma$ point-source magnitude limit $i_{\rm AB} \sim 24.1$, and a mean seeing of 0.6". The effective area is 129.2 sq. degrees after masking out bright stars and other image artifacts (L17). We use a new version 3.0 of the shape catalogue, with shapes measured with \textsc{lensfit} down to magnitude $i_{\rm AB} < 24.7$. This new version benefits from internal calibration in \textsc{lensfit} based on image simulations inherited from the CFHTLenS project. Shape measurements are accurate at the 2\% level, without relying on any additional linear correction. In addition, this new catalogue contains about 40\% more galaxies, mostly because of a better handling of galaxy de-blending and instrument artefacts in \textsc{lensfit} (priv. com. with L. van Waerbeke). 

Photometric redshifts in the original version of the catalogue \citep{bundy2015} were computed with BPZ \citep{benitez2000} using {\it ugriz} from the Stripe 82 co-adds \citep{annis2014} and $UJHK$ from UKIDSS. We use nearest-neighbor interpolation in sky coordinates, $i$ magnitude and {\it g-r, r-i, i-z} color space to get photometric redshifts for the new galaxies. We verify that the redshift distribution is unchanged. We apply the same procedure as in the CFHTLS fields to estimate the bias due to photometric redshifts in our GGL measurements. However, given the relatively shallow spectroscopic survey coverage of the CS82 field compared to CFHTLS fields (90\% completeness reached at $i_{\rm AB} = 22.5$),  we are forced to select galaxies only down to $i_{\rm AB} < 22.5$. \ch{Although this cut is quite severe, it allow us to confidently model and correct photometric redshift bias in this field. The lack of deeper spectroscopic information prevents us from exploiting the complete weak-lensing catalogue}. For $i_{\rm AB} < 22.5$ sources and CMASS lenses, we find a bias $b_z = -0.028\pm0.006$. In contrast to L17, we apply no cut based on the \textsc{odd} quality flag, because we find it has no impact on our lensing measurements given our stringent cut in magnitude. Our final catalogue contains 2.2 million galaxies. The galaxy density is $n_{\rm eff} = 4.7$ galaxies arcmin$^{-2}$. The median redshift is $z_{\rm median} = 0.53$.

\subsection{Spectroscopic dataset: the BOSS CMASS sample}

The BOSS spectroscopic survey \citep{eisenstein2011} is a program of the SDSS project \citep{gunn2006}. The Constant (Stellar) Mass (CMASS) galaxy sample is one of the galaxy samples observed in this survey. It consists of galaxies selected with the SDSS photometry, such that they lie in the redshift range $0.43< z < 0.7$, and represent a sample of galaxies approximately volume-limited in stellar mass \citep{reid2016}. Early clustering analysis found that CMASS galaxies lie in massive haloes, with mean halo mass of $2.6\times10^{13} h^{-1} {\rm M}\odot$, a large scale bias of $b \sim 2.0$ and a satellite fraction of 10\% \citep{white2011}.

\begin{figure}
\centering
\includegraphics[width=\linewidth]{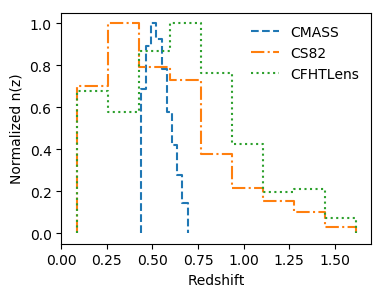} 
\caption{Redshift distribution of CMASS galaxies (blue) compared to CS82 (orange) and CFHTLens (green) source distributions, after weak-lensing selection has been performed. Weak-lensing $n(z)$ are based on photometric redshifts (see text for details).}
\label{fig:nz}
\end{figure}

We use the public DR12v5 version of the CMASS catalogue \citep{alam2015}. The galaxy surface density is about 100 deg$^{-2}$ \citep{reid2016}. We only consider CMASS overlapping with our 4 lensing fields, i.e. covering an area 250 sq. degrees. Our catalogue of lenses contains 28,039 CMASS galaxies, distributed as reported in Table~\ref{tab:cmass}. The redshift distribution of CMASS galaxies compared to CS82 and CFHTLens lensing sources is shown in Figure~\ref{fig:nz}. 

\begin{table*}
\centering
\caption{Number of CMASS galaxies per field, effective lensing area after masking and number of weak-lensing sources.}
\begin{tabular}{cccccc}
\hline
Field & \# CMASS & SDSS area  &  SDSS field size  & Eff. area  & \# sources \\
&& [deg$^2$]&  [deg$\times$deg] & [deg$^2$] & [$\times10^6$] \\
\hline
S82 & 18,675 & 219.8&   $87.6\times2.51$ & 129.2 & 2.19\\. 
W1 & 3,924 &  54.14& $8.66\times6.3$ & 63.8 & 1.66\\ 
W3 & 3,694 & 41.91&$11.5\times6.6$ & 44.2 & 1.26\\ 
W4 & 1,746 & 22.16& $5.7\times5.61$ & 23.3 & 0.62\\ 
\hline
\end{tabular}
\label{tab:cmass}
\end{table*}

In spite of a careful photometric selection, the observed CMASS galaxy sample remains contaminated by various observational effects \citep{ross2012}. We take them into account by applying the galaxy weights $w_g = w_{star} w_{see} (w_{zf} + w_{cp} -1)$ as defined in \citep{ross2017}. We also include the FKP \citep{feldman1994} weights with the parameter $P_0=20,000 h^{-3}{\rm Mpc}^3$ \citep{ross2012}, such that the noise in the power spectrum is minimum at the BAO scale $k = 0.1 h\ {\rm Mpc}^{-1}$. Although not optimal for our study focused on small scale clustering, this value of $P_0$ allows consistent comparison with previous measurements. For consistency, we take the same value of $P_0$ for our mock catalogues and data. Finally, we use the \verb?DR12v5_random0? catalogues trimmed to the regions overlapping with WL data.

\section{Measurement estimators}

\subsection{Galaxy-galaxy lensing estimation}

We compute $\Delta \Sigma_{\rm gm}(r_{\rm p})$ using the estimator

\begin{multline}
\Delta \Sigma_{\rm gm}(r_{\rm p}) = \frac{\sum_{\rm l,s}^{N_{\rm l,s}} \Sigma_{\rm cr}(z_{\rm l}, z_{\rm s}) w_{\rm l,s} \epsilon_+}{\sum_{\rm l,s}^{N_{\rm l,s}} w_{\rm l,s}} \\ - \frac{\sum_{\rm r,s}^{N_{\rm r,s}} \Sigma_{\rm cr}(z_{\rm r},z_{\rm s}) w_{\rm r,s} \epsilon_+}{\sum_{\rm r,s}^{N_{\rm r,s}} w_{\rm r,s}}
\label{eq:estggl}
\end{multline}

\noindent where the summation runs over all pairs of sources and lenses at redshifts $z_{\rm l}$ and $z_{\rm s}$ separated by the projected radius $r_{\rm p}$ to within a given bin width. The subscript "r" denotes the random catalogue of lensing objects. Our number of random objects $N_{\rm r}$ is 10 times the number of lenses $N_{\rm l}$. Their redshift distribution $n(z)$ is the one from CMASS galaxies \citep{nuza2012}. The subtraction of the random signal decreases the variance at large scales \citep{singh2017,shirasaki2017}. $\epsilon_+$ represents the tangential component of a source ellipticity around a lens. The weight $w_{\rm l,s} = \Sigma_{\rm cr}^{-2}(z_{\rm l},z_{\rm s}) w_{\rm s} $ is the product of the shape measurement weight $w_{\rm s}$ from \textsc{lensfit} and the critical density. This inverse variance scheme downweights pairs which are close in redshift \citep{mandelbaum2006}. The critical lensing density $\Sigma_{\rm cr}(z_{\rm l},z_{\rm s})$ in comoving units is defined as 

\begin{equation}
\Sigma_{\rm cr}(z_{\rm l},z_{\rm s}) = \frac{c^2}{4\pi G (1+z_s)^2} \frac{D_S}{D_{LS} D_{L}}
\end{equation}

\noindent where $D_S, D_{LS}, D_L$ are the observer-source, lens-source and observer-lens angular diameter  distances \footnote{The factor $(1+z_s)^2$ is missed in Eq. 10 of \citet{delatorre2017}, but was properly taken into account in the calculations.} We use the best-fit estimate of the photometric redshift to compute the distances, instead of the full probability distribution, as suggested in \cite{blake2016}. However, our approach described below and based on full ray-tracing simulations consistently takes this simplification into account.

\subsection{Anisotropic galaxy clustering estimation}
\label{sec:estimators}

We compute the two-point galaxy correlation function in the polar and \ch{cartesian coordinate systems}. The anisotropy in the signal is due to the RSD effect we are after. \ch{The estimator is the same in each coordinate system and is defined as
\begin{equation}
\xi(x,y) = \frac{GG(x,y) - 2GR(x,y) + RR(x,y)}{RR(x,y)}
\label{eq:estrsd}
\end{equation}
\noindent where $(x,y)=(s,\mu)$ or $(r_p,\pi)$. $GG$, $GR$ and $RR$ are respectively the normalized number of pairs between galaxy-galaxy, galaxy-random, and random-random at a given separation. }

We compress the information contained in $\xi(s,\mu)$ by projecting it on the Legendre polynomials using the expressions for the correlation-function multipole moments

\begin{equation}
\xi_\ell(s) = \frac{2\ell +1}{2} \int_{-1}^{1}\xi(s,\mu) L_\ell(\mu) \d\mu\,
\end{equation}

\noindent where $L_\ell$ is the Legendre polynomial of order $\ell$. We use the monopole and quadrupole $\ell=(0,2)$ only because, the higher order non-null multipoles are too noisy.

We also compute the projected correlation function $w_{\rm p}(r_{\rm p})$ \st{from the anisotropic correlation function} by projecting $\xi(r_p,\pi)$ along the line-of-sight, such that

\begin{equation}
w_{\rm p}(r_{\rm p}) = 2\int_0^{\pi_{\rm max}} \xi(r_{\rm p},\pi) \d \pi\,,
\end{equation}

\noindent where we find the optimal value $\pi_{\rm max} = 40h^{-1}\ {\rm Mpc}$ to minimize the noise due to the limited number of pairs in our fields. 

\subsection{Joint lensing and clustering likelihood}

We perform a maximum likelihood analysis to derive the cosmological parameters from the GGL and RSD measurements. \ch{In each field $i$, we measure the data vector ${\bf d}^i = (\xi_0^i, \xi_2^i, \Upsilon_{\rm gm}^i)$} and we compute the likelihood function per field $\mathcal{L}^i$ such that
\begin{equation}
-2 \ln \mathcal{L}^i = ({\bf d}^i - {\bf m})^T \hat{\bf \Phi}^i ({\bf d}^i - {\bf m}),
\end{equation}

\noindent where $\bf m$ is the model prediction, and $\hat{\bf \Phi}^i$ is the precision matrix estimated from the simulations. 

Our 4 fields are statistically uncorrelated, and therefore the global likelihood is just the product of the individual likelihoods for each field
\begin{equation}
\mathcal{L}_{\rm tot} = \mathcal{L}_{\rm CS82} \times \mathcal{L}_{\rm W1} \times \mathcal{L}_{\rm W3} \times \mathcal{L}_{\rm W4}
\end{equation}

Field W4 partly overlaps with field S82, but this overlapping represents $<$6\% of the total area. In addition, CFHTLens catalogue used for W4 goes deeper than CS82 catalogue used for S82, thus decreasing further the correlation between the 2 fields.

\section{Simulations}
\label{sec:simus}

\subsection{Lightcones and lensing mock catalogues}
\label{sec:simu}

In order to accurately estimate large scale variance and possibly unveil new systematic errors, we produce light-cones with the same geometry as the observed fields. We use the BigMultidark N-body simulation, as it appears to be a good compromise between particle resolution and cosmological volume  \citep[$m_{\rm p} = 2.5\times10^{10}\ h^{-1}\ M_\odot$, $L_{\rm box} = 2.5 h^{-1} \rm{Gpc}$, Planck cosmology with $h=0.6777$]{klypin2016}. Following the approach described in \cite{giocoli2016}, we simulate 4 fields CS82, W1, W3 and W4, with lightcones extending up to redshift $z=2.3$ for the CFHT-LS fields, and $z=2$ for CS82. Lensing properties, such as deflected positions, shear and convergence, are computed by ray-tracing through 25 lens planes separated by $161h^{-1}$\ Mpc comoving \citep{giocoli2016} using the GLAMER code \citep{metcalf2014,petkova2014}. The spatial resolution of the lensing maps is 6 arcseconds. 

\subsubsection{Lensing properties}

We simulate lensing catalogues of sources including survey mask, intrinsic shape and photometric redshift noises. 
For survey mask, we simulate lensing sources at the location of the observed weak-lensing sources. \ch{Thus, we naturally reproduce the footprint, as well as the holes around bright stars and other artefacts in the real weak-lensing catalogue. Effects due to the intrinsic clustering of sources in projection are also included.} 
For the intrinsic ellipticities of the sources, \ch{we randomly draw observed ellipticites $\epsilon^{\rm obs}$ from the weak-lensing catalogue, that we multiply by a random orientation $\phi^{\rm int}$, such that $\epsilon_1^{\rm int} = \epsilon^{\rm obs}\cos(2\phi^{\rm int})$ and $\epsilon_2^{\rm int} = \epsilon^{\rm obs}\sin(2\phi^{\rm int})$. } \st{we randomly draw values from \textsc{lensfit} ellipticity distribution. }

\subsubsection{Photometric redshifts}

To simulate photometric redshifts with catastrophic failures, \ch{we design a method related to the one described in \citet{lima2008}, also referred as the DIR method in the KiDS survey \citet{hildebrandt2017}.  We start by estimating the true redshift distribution $n_{\rm true}(z)$ for our CFHTLens ($i_{\rm AB} < 24$) and CS82 ($i_{AB} < 22.5$) weak-lensing catalogues from our spectroscopic calibration sample described in Appendix~\ref{sec:zphbias}. 
In practice, we compute the histograms of the weak-lensing (WL) and spectroscopic (ZP) catalogues in the magnitude-color space $(i,g-r,r-i,i-z)$, that we limit to the region $([18;25],[-1;3],[-1;3],[-1;3])$. We apply the same binning for both catalogues. For each bin of coordinates ${\bf m}$, we derive the weights $W({\bf m}) = N_{WL}({\bf m}) / N_{ZS} ({\bf m})$, where $N$ is the number of sources per bin. We assume all sources in a bin have the same weight. Finally, we obtain the true  distribution in redshift bin $i$ with the following sum

\begin{equation}
n_{\rm true}(z_i) = \int \d {\bf m} N_{ZS}(z_i|{\bf m}) W({\bf m})\;.
\end{equation}

\noindent A drawback of this approach is that if spectroscopic selection does not cover part of the redshift range, then it truncates  $n_{\rm true}(z)$. However, we see in the following that the coverage is sufficient for our purpose.}

\begin{figure}
\centering
\includegraphics[width=\linewidth]{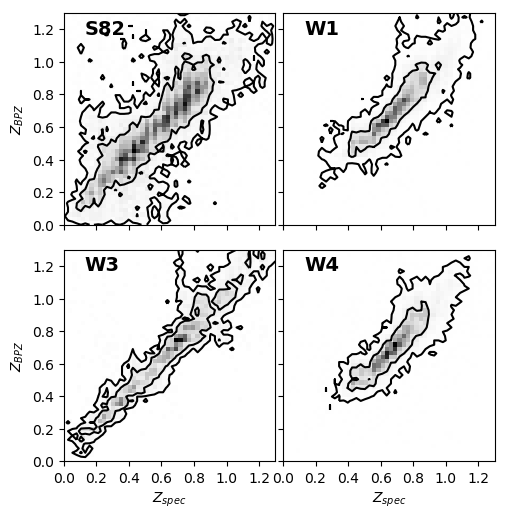}
\caption{Probability distribution of having a photometric redshift with BPZ \st{, given} \ch{and} a spectroscopic redshift {for each field}. Contours are given for 1,2 and 3$\sigma$ C.L.}
\label{fig:pz2d}
\end{figure}

 \st{assign true redshifts by randomly drawing redshifts from the redshift distributions $n(z)$ of the CFHTLens and CS82 sources}. We  compute \st{photometric redshifts by sampling} the joint probability $P(z_{\rm BPZ},z_{\rm spec})$ for each field, as shown in Figure~\ref{fig:pz2d}. \ch{We observe} that the spectroscopic redshift completeness at $z<0.5$ in field W1 and W4 is very low, because most of the redshifts come from the CMASS sample. Fortunately, this has little impact on our simulation of photometric redshift noise, because our analysis focuses on the cross-correlation of CMASS galaxies with lensing sources at $z>0.5$. \ch{We also observe} that the scatter in the $z_{\rm BPZ}$ of CS82 field is almost twice as large as in field W3, \ch{and differs between the 3 CFHTLens fields. This justifies our field-by-field treatment of the photometric redshift noise. Finally, we assign photometric redshifts to the simulated sources by randomly drawing a photometric redshift from $P(z_{\rm BPZ},z_{\rm spec})$, where we assume the spectroscopic redshift $z_{\rm spec}$ is the true redshift, assigned at the beginning of the procedure. }

\subsection{Spectroscopic CMASS mock catalogues}

\begin{figure}
\begin{tabular}{cc}
\includegraphics[width=0.5\linewidth]{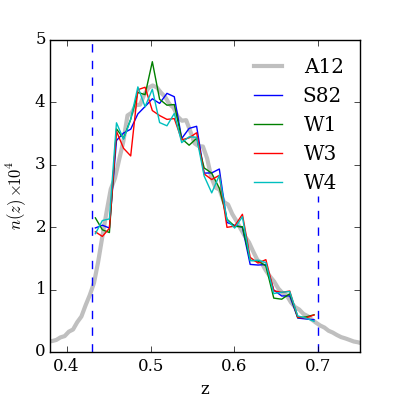} &
\includegraphics[width=0.5\linewidth]{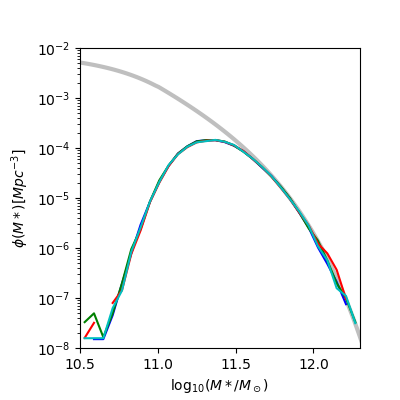} 
\end{tabular}
\caption{{\bf left panel:} Number density of CMASS mock galaxies for our 4 simulated fields. Limits of our analysis are marked with blue dashed lines. Measurements from \cite[A12]{anderson2012} are in gray. {\bf right panel:} CMASS stellar mass function for the 4 fields reproducing the observed incompleteness. The mock catalogue is complete at high mass in agreement with the model proposed in \cite{rodriguez2016} in gray.}
\label{fig:nzcmass}
\end{figure}

We adopt the Stellar to Halo Abundance Matching (SHAM) procedure described in \cite{rodriguez2016} to produce CMASS mock catalogues. Starting from the \textsc{Rockstar} public catalogues \citep{behroozi2013}\footnote{https://www.cosmosim.org/cms/simulations/bigmdpl/}, we compute a scattered peak velocity $V_{\rm peak}^{\rm scat} = (1+\mathcal{N}(0, \sigma_{\rm SHAM}))V_{\rm peak}$, where $\mathcal{N}$ is the Normal distribution, and $\sigma_{\rm \ch{S}HAM} = 0.31$.  We also simulate the CMASS incompleteness in stellar mass and redshift, based on the Stellar Mass Function (SMF) from the Portsmouth \textsc{sed-fit} DR12 stellar mass catalogue with Kroupa initial mass function \citep{maraston2013}. We bin the catalogue in 12 redshift intervals between $0.43 < z < 0.7$  and in 18 stellar mass bins between $10.5 < log_{10}(M*/M_\odot) < 12.3$. Thus, we obtain a tabulated SMF that we can interpolate in stellar mass and redshift. Finally from cumulative stellar mass and halo mass functions, we construct a number density matching such that
$n_{\rm gal}(>M_*^i) = n_{\rm halo}(>V_{{\rm peak},i}^{\rm scat})$. Since different cosmologies were assumed in the Portsmouth catalogue and in the BigMultidark simulations, \ch{$h=0.73$ and $h=0.6777$ respectively} , \st{we consistently take into account in} \ch{we renormalized} the 
stellar masses \ch{to the BigMultidark cosmology} \st{of the variation from $h=0.73$ to $h=0.6777$}. As shown in Figure~\ref{fig:nzcmass}, our number densities for each of the 4 fields are in good agreement with the measurements from \cite{anderson2012}.  

We also include the effect of peculiar velocities by summing together in redshift-space the halo position ${\bf r}_c$ and the peculiar velocity vector ${\bf v}$ in real space using the relation 
${\bf s} = {\bf r}_c + \frac{{\bf v} \cdot {\bf {\hat r}}}{a\,H(z_{\rm real})}$, where $\bf{\hat{r}}$ is the line of sight unit vector, $a$ is the scale factor,  and $H(z_{\rm real})$ is the Hubble parameter at redshift $z_{\rm real}$, the redshift corresponding to $\bf{r}_c$. Finally, we mask the borders of the square simulated fields W3 and W4 to reproduce their complex geometry, and we compute the FKP weights, assuming  the same $P_0 = 20,000 h^{-3} Mpc^3$ as in the data. Since the data are corrected for fiber collision, redshift failure, stellar density and seeing, we do not simulate these effects.

\subsection{Bias due to photometric redshift noise}

\begin{figure}
\centering
\includegraphics[width=\linewidth]{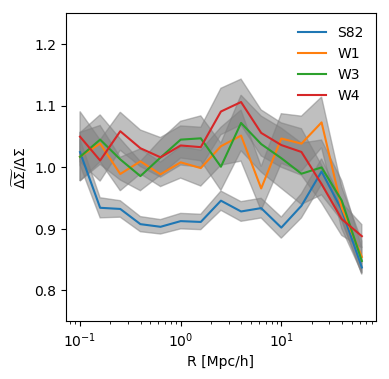}
\caption{Comparison of lensing measurements performed on simulated catalogues affected and not affected by photometric redshifts noise $\widetilde {\Delta \Sigma}$ and $\Delta \Sigma$ respectively. Grey shaded areas correspond to the uncertainties on the mean value obtained by resampling the multiple noises in different light cones.}
\label{fig:dsig_cv}
\end{figure}

We compute successively the lensing signal for catalogues with and without photometric redshift noise, and compare the measurements in Figure~\ref{fig:dsig_cv}. We find that the large photometric scatter observed in field S82 (Figure~\ref{fig:pz2d}) seems to result in a bias of about 10\% in the lensing signal at scales $R< 10h^{-1} {\rm Mpc}$, whereas the CFHTLens fields seem insignificantly affected. We argue that this might explain the discrepancy highlighted in L17 between lensing measurements obtained with real and mock data. Indeed, in the following, we show that our lensing measurements with mock data contaminated by photometric redshift noise are in agreement with real data. 

\subsection{Bias due to small scale modeling}

\begin{figure}
\centering
\includegraphics[width=\linewidth]{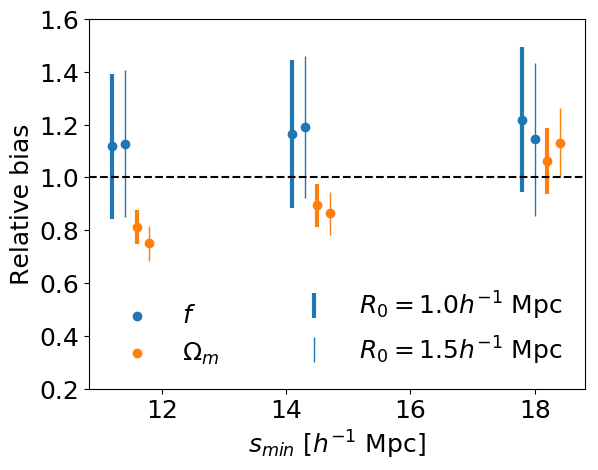} 
\caption{Bias between recovered parameters  $f$ and $\Omega_{\rm m}$ relative to the values used in the mocks, as a function of the minimum scale of the multipoles $s_{\rm min}$, and the  cut-off radius of the $\Upsilon(R,R_0)$ lensing estimator. Values of $s_{\rm min}=17.8\ h^{-1} {\rm Mpc}$ and $R_0 = 1.0h^{-1} {\rm Mpc}$ provide unbiased estimates of $f$ and $\Omega_{\rm m}$. All measurements were performed without tapering smoothing in the covariance matrices. }
\label{fig:smin}
\end{figure}

We use the simulation to quantify the bias in the estimation of the cosmological parameters $f$ and $\Omega_{\rm m}$ due to our model prediction of the small scales. Successively, we cut data points of $\xi_0$ and $\xi_2$ at scales $s_{\rm min} = 11.2$, 14.1 and $17.8\ h^{-1}\ {\rm Mpc}$, and $\Upsilon_{\rm gm}$ at scales $R_0 = 1.0$ and $1.5\ h^{-1}\ {\rm Mpc}$. Overall, we find that the values $s_{\rm min} = 17.8\ h^{-1}\ {\rm Mpc}$ and $R_0 = 1\ h^{-1}\ {\rm Mpc}$ provide the best compromise between systematic bias and statistical precision as can be seen in Figure~\ref{fig:smin}.

\subsection{Covariance matrices}

\begin{table*}
\caption{Properties of the simulated fields in terms of independent mock catalogue, random resampling of lensing shape noise and photometric redshifts per catalogue, and number of sub-regions for Jackknife resampling. }
\label{tab:simfields}
\centering
\begin{tabular}{ccccc}
\hline
Field Name & Size [deg] & \parbox[t]{6em}{\centering Number \\ of mocks} & \parbox[t]{7em}{\centering Number \\ of realisations} & \parbox[t]{5em}{\centering Number of subregions} \\
\hline
CS82 & $87.15\times2.58$ & 4 & 60 & 16 \\
W1 & $8.7\times6.3$ & 15 & 4 & 12\\
W3 & $11.7\times6.6$ & 11 & 4 & 16 \\
W4 & $5.7\times5.6$ & 27 & 3 & 9\\
\hline
\end{tabular}
\end{table*}

In order to obtain an unbiased estimate of the precision matrices, we need minimal errors in the covariance matrices, and therefore a large number of mock catalogues. \ch{Noise in the covariance matrices increases the errors on the model parameter estimation \citep[see e.g.][]{taylor2014}}. Unfortunately, we are limited by the size of our simulation box $L = 2h^{-1} {\rm Gpc}$. \cite{escoffier2016} proposed a method to increase the number of mocks, based on Jackknife resampling of the mock catalogues (see Table~\ref{tab:simfields}).  Following their prescription, we split each catalogue into $N_{\rm JK}$ spatial subregions, and measure the clustering and lensing signals in each Jackknife subsample using estimators given in Eq.~\ref{eq:estrsd} and Eq.~\ref{eq:estggl}. The covariance matrix for each mock catalogue is then

\begin{equation}
^{(m)}\hat{C}_{ij}^{\rm JK} = \frac{N_{\rm JK}-1}{N_{\rm JK}} \sum_{k=1}^{N_{\rm JK}} (d^k_i - \bar{d}_i)(d^k_j - \bar{d}_j)\,,
\end{equation}

\noindent where the mean vector is obtained from the Jackknife samples
\begin{equation}
\bar{d}_j = \frac{1}{N_{\rm JK}} \sum_{k=1}^{N_{\rm JK}} d_i^k\,.
\end{equation}

In addition, given our limited number of independent mock catalogue $N_{\rm m}$, we increase their number for lensing by resampling $N_{\rm r}$ times the observed lensing ellipticity distribution function, and the photometric redshifts distribution. We find this strategy to efficiently improve the accuracy of the covariance matrix for the lensing, especially at small scales. The final covariance matrix is therefore obtained by averaging the Jackknife covariance matrices

\begin{equation}
\bar{C}_{ij} = \frac{1}{N_{\rm m}\times N_{\rm r}} \sum_{m=1}^{N_{\rm m}\times N_{\rm r}} {}^{(m)}\hat{C}_{ij}^{\rm JK}
\end{equation}

Finally, we compute the precision matrix 

\begin{equation}
\hat{\Phi}_{ij} = [\bar{C}_{ij}]^{-1}\,.
\end{equation} 

\cite{escoffier2016} have shown that this expression provides an unbiased estimate of the true precision matrix. 

\begin{figure}
\centering
\includegraphics[width=\linewidth]{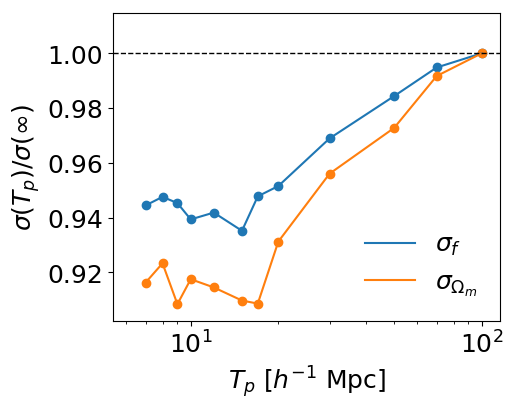} 
\caption{Variation of the relative errors on the parameters $f$ and $\Omega_{\rm m}$, as a function of the 
smooth scale $T_{\rm p}$ in the covariance matrices. There is no improvement below $T_{\rm p} = 12\ h^{-1}$}  

\label{fig:ts}
\end{figure}

\begin{figure}
\centering
\begin{tabular}{c}
\includegraphics[width=\linewidth]{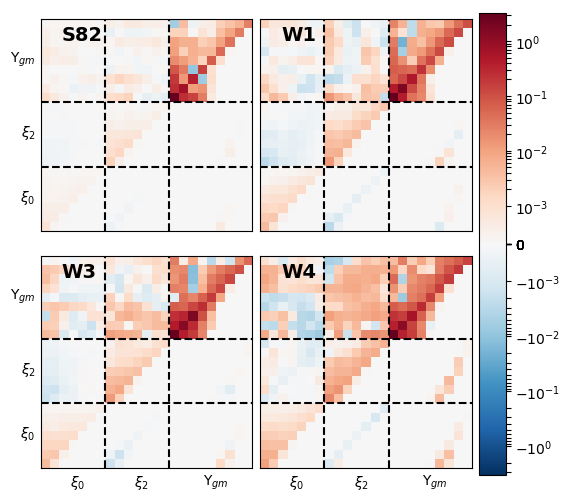} \\
\includegraphics[width=\linewidth]{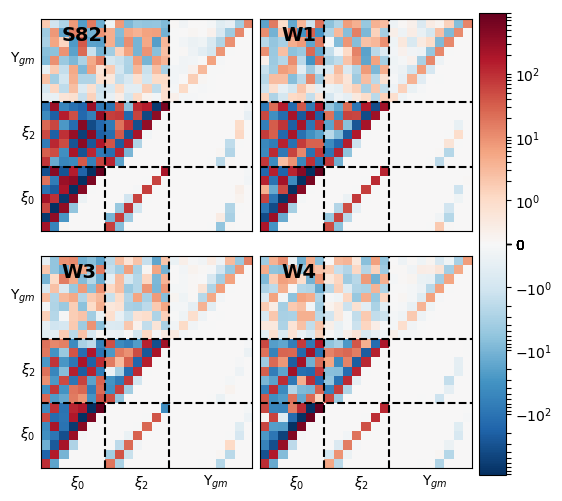}
\end{tabular}
\caption{Matrices of covariance (left panel) and precision (right panel) in logarithmic color scale for the 4 fields used in this analysis. In both panels, the upper triangular part of the matrices represents the case without tapering, while the lower part represents the case with tapering $T_{\rm p}=12h^{-1}\ {\rm Mpc}$. \ch{Noise between far apart scales is significantly decreased and the errors on the model parameters converge to a minimum.} }
\label{fig:cov}
\end{figure}

In spite of our resampling strategy, our covariance matrices are still noisy. Therefore, we adopt the tapering method proposed by \cite{paz2015} to damp the noise by a filter function beyond a given tapering scale $T_{\rm p}$. This technique is based on the assumption that correlation between \ch{pairs of data points}  far apart is negligible and little information is lost by treating these points as being independent. \ch{Although very efficient, it is commonly accepted that this method might inadvertently remove non-Gaussian terms \citep{paz2015}. However this effect is beyond the scope of this analysis given our data and the range of scales investigated here. In Fig.~\ref{fig:ts}, we observe that large tapering yields errors similar to no tapering. On the opposite, small tapering zeros all off-diagonal terms, and can also lead to overestimated errors. We find the errors on $f$ and $\Omega_{\rm m}$ to reach a minimum value at} the tapering scale $T_{\rm p} \sim 12 h^{-1} {\rm Mpc}$. We adopt this scale in the rest of this analysis. We should note that all measurements were performed with $s_{\rm min} = 14.1h^{-1} {\rm Mpc}$ and $R_0 = 1.5h^{-1} {\rm Mpc}$. However, we repeated some measurements with our final setup ($s_{\rm min} = 17.8h^{-1} {\rm Mpc}$ and $R_0 = 1.0h^{-1} {\rm Mpc}$) and found that these parameters have almost no impact on the tapering scale behavior. The covariance and precision matrices obtained before and after tapering at this scale are shown in Fig.\ref{fig:cov}. \ch{ We can observe that the noise in the off-diagonal terms is significantly reduced after tapering. This is particularly obvious between clustering and lensing, which cover very different range of scales.}

\section{Cosmological results}
\label{sec:results}

The quality tests and errors assessment that we performed with the simulations give us confidence that our dataset can lead to \st{interesting} \ch{reliable} cosmological constraints. 

\subsection{galaxy-clustering and galaxy-galaxy lensing measurements}

\begin{figure}
\centering
\includegraphics[width=\linewidth]{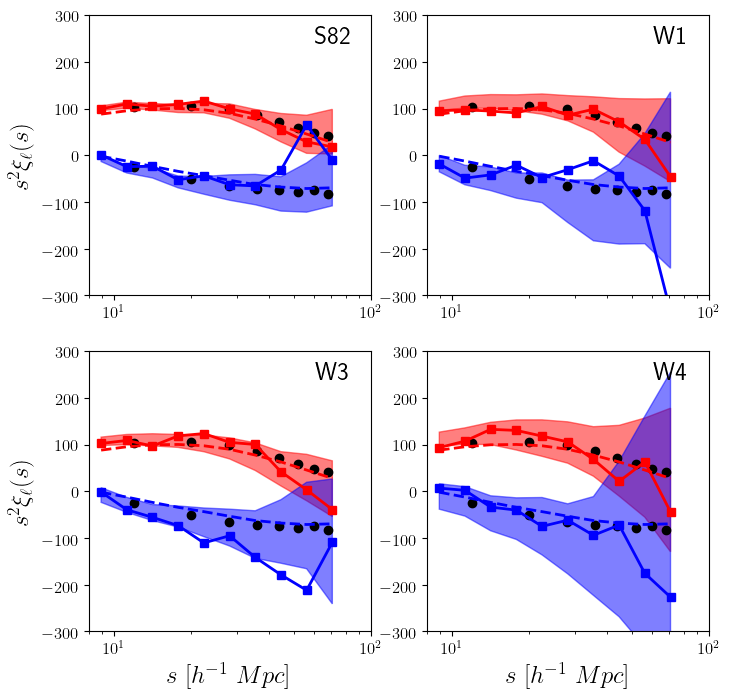} 
\caption{Monopole (red) and quadrupole (blue) measurements with mock catalogs (shaded region), real data (solid lines) and theoretical predictions with a linear bias parameter $b_1=1.8$ (dashed lines). Black dots represent pre-reconstruction measurements with the full DR12v5 CMASS sample from \cite{cuesta2016}. }
\label{fig:mock_xi}
\end{figure}

\begin{figure}
\centering
\includegraphics[width=\linewidth]{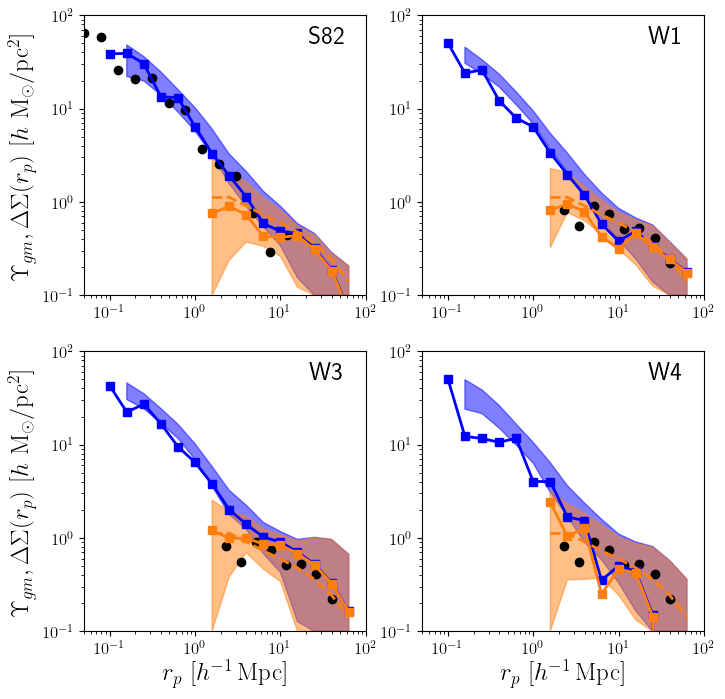}
\caption{Filtered $\Upsilon_{\rm gm}$ and non-filtered $\Delta \Sigma$ GGL measurements with mocks (shaded regions), $\Delta \Sigma$ and $\Upsilon$ data (blue and cyan points respectively), and theory with a linear bias parameter $b_1=1.8$  (dashed line). Black dots in S82 panel represent $\Delta \Sigma$ measurements from L16, and $\Upsilon_{\rm gm}$ measurements from \cite{alam2016} in CFHTLS panels. }
\label{fig:mock_dsig}
\end{figure}

In Figure \ref{fig:mock_xi} \ch{and \ref{fig:mock_dsig}}, we show our RSD and GGL measurements, along with our theoretical predictions, assuming the fiducial parameters of the simulation, and a constant linear bias $b_1 = 1.8$. We find a good agreement within 1$\sigma$ C.L. between mocks, data and theoretical predictions for all fields. \ch{We notice that the quadrupole of the correlation function measurement in the field W3 is lower than the $1\sigma$ C.L.,  and that the GGL measurement in the field W4 is lower than $1\sigma$ C.L. at scales $R < 1h^{-1}\ {\rm Mpc}$ . For field W3, we found that setting $\sigma_8(z=0.57) = 0.9$ and $b_1=1.5$ could reconcile predictions with measurements, thus suggesting a sample variance effect. These values are within the 3$\sigma$ C.L. of the RSD-only fit of the data (see~\ref{fig:om_f}). For field W4, we attribute the discrepancy to our poor modeling of baryonic or lensing effects at small scales}, that average out too slowly in the data to reproduce the simulated dark-matter only profile. Nonetheless, the overall good agreement \st{we find, especially for field S82,} gives us confidence that we can proceed with the cosmological analysis.

\subsection{Growth of structure and background constraints}

\begin{figure}
\centering
\includegraphics[width=\linewidth]{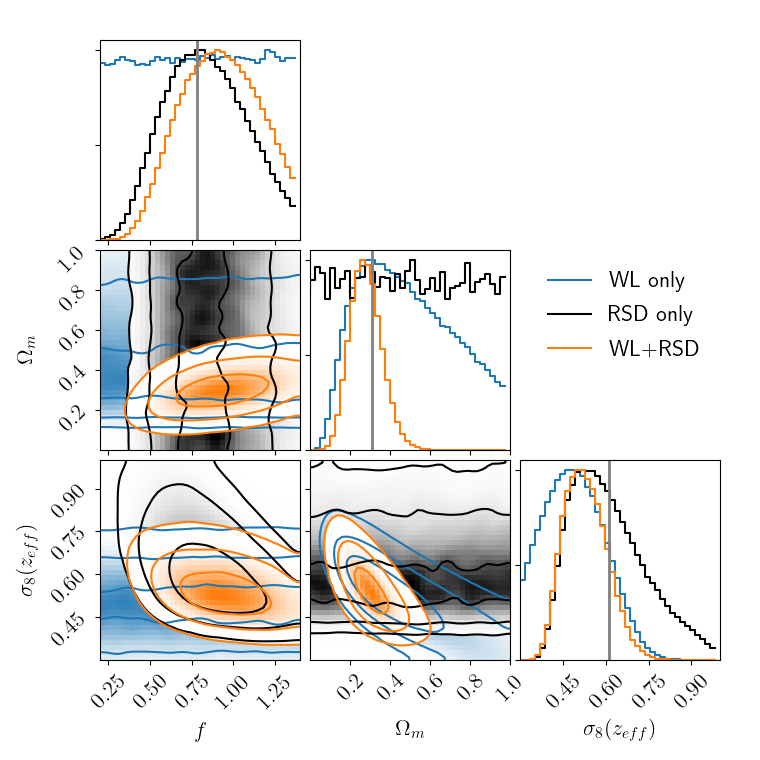} 
\caption{Improvement on estimating $\Omega_{\rm m}$ and $\sigma_8$ when combining RSD and WL measurements. Blue and black curves are respectively obtained with WL and RSD constraints only. Orange curves are obtained with the combination of WL and RSD. \ch{Contours are given at 1, 2 and 3$\sigma$ C.L.} Vertical lines indicate Planck TT,TE,EE+lowE 2018 results. }
\label{fig:om_f}
\end{figure}

We estimate the cosmological parameters $f$, $\sigma_8$ and $\Omega_{\rm m}$ by combining $\xi_0$, $\xi_2$ and $\Upsilon_{\rm gm}$ measurements. The power of this combination to break the degeneracy between $f$ and $\sigma_8$ has already been demonstrated \citep[see e.g. DLT17,][]{joudaki2017}. In this paper, we move one step further by estimating $\Omega_{\rm m}$ as well from the data. Figure~\ref{fig:om_f} shows the independent lensing, clustering and combined constraints on these parameters. Best fit values and $1\sigma$ error estimates are reported in Table~\ref{tab:fitresults}. A corner plot with all the parameters involved in the fit is reported in the Appendix in Figure~\ref{fig:corner}. On the one hand, we find that GGL alone constrains $\Omega_{\rm m}$ at 45\% and $\sigma_8$ at 22\%. It provides no constrain on the structure growth rate $f$. On the other hand, RSD also constrains $\sigma_8$ at 20\% but leaves $\Omega_{\rm m}$ completely unconstrained as expected from the model.
When used in combination, GGL and RSD measurements yield 12\% precision constraint on $\sigma_8$, i.e. almost as if the 2 datasets were independent. In fact, Figure~\ref{fig:om_f} shows that the well-known WL degeneracy between $\Omega_{\rm m}$ and $\sigma_8$ intersects almost perpendicularly with the constraint on $\sigma_8$ from RSD.  

\begin{figure}
\centering
\includegraphics[width=\linewidth]{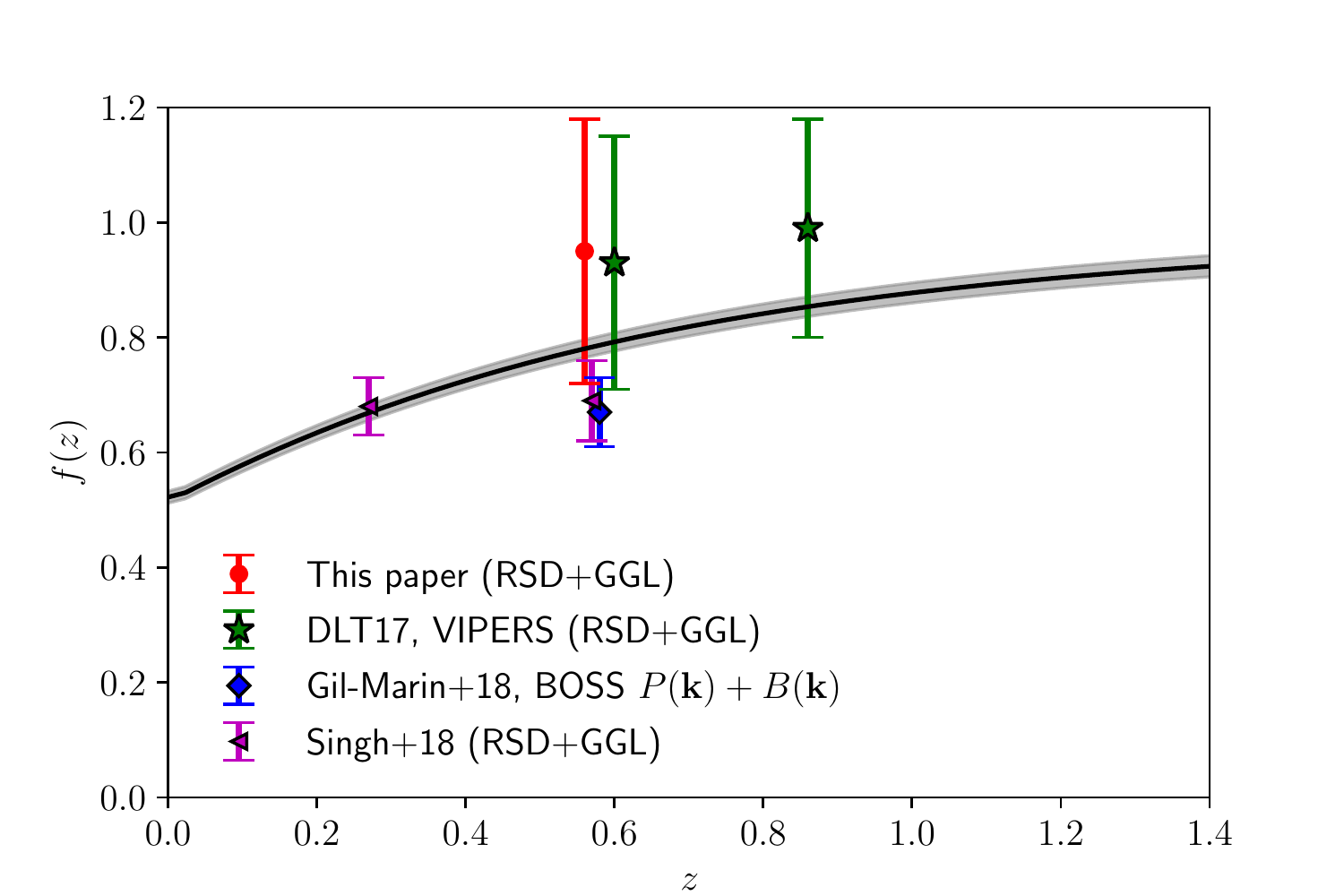}
\caption{Growth rate $f$ as a function of redshift compared to recent measurements. The black line and surrounding grey shared area indicate the Planck TT,TE,EE+lowE 2018 mean and $1\sigma$ uncertainty predictions for $\Lambda$CDM-GR flat model.}
\label{fig:fz}
\end{figure}

In Figure~\ref{fig:fz}, we present our estimate of the growth rate $f$, and compare to other measurements. In spite of having a wider area, we obtain a constraint similar to the one found in DLT17 with VIPERS. Clearly, the number of RSD tracers determines the precision. In both analysis, we have about 28,000 galaxies in the range $0.5<z<0.7$. Regarding weak-lensing, the number densities of background sources at $z>0.7$ in both analysis are similar. We have $n_{\rm eff} = 3.45$ in CFHTLS fields and $n_{\rm eff} = 2.33$ in CS82 and CFHTLS fields combined. 

We also compare our results with analyses performed on the full CMASS sample. \cite{singh2018} performed a joint analysis with Planck CMB lensing and SDSS galaxy lensing and obtained three times tighter constraints than ours. Their results are in agreement with ours at the 1$\sigma$ C.L.

Finally, combining CMASS power-spectrum and bi-spectrum, \cite{gilmarin2017}  also obtained very competitive constraints at redshift $z=0.57$ in agreement with ours. These two estimates find a tension on $f$ with Planck predictions at $z=0.57$. Interestingly, this tension was also observed in other RSD analysis with the CMASS sample, but not with the LOWZ sample \citep[e.g.][]{alam2016,beutler2014}.

\begin{table*}
\centering
\begin{tabular}{cccc}
\hline
Parameters & RSD only & GGL only & GGL+RSD  \\
\hline
$\alpha$ & $1.02\pm0.05$ & $1.01\pm0.05$ & $1.02\pm0.05$ \\
$\epsilon$ & $0.01\pm0.05$ & $-0.01\pm0.06$ & $0.00\pm0.05$ \\
$f(z=0.57)$ & $0.86\pm0.24$ & --- & $0.95\pm0.23$ \\
$\sigma_8(z=0.57)$ & $0.63\pm0.13$ & $0.50\pm0.11$ & $0.55\pm0.07$ \\
$\Omega_{\rm m}$ & --- & $0.51\pm0.23$ & $0.31\pm0.08$ \\
\hline
$b_1$ & $2.09\pm0.43$ & $1.94\pm0.55$ & $2.33\pm0.33$ \\
$b_2$ & $-0.06\pm0.53$ & $0.03\pm0.54$ & $-0.05\pm0.53$ \\
$\sigma_v$ & $4.55\pm1.68$ & --- & $4.20\pm1.64$ \\
\hline
$S_8 = \sigma_8\sqrt{\Omega_{\rm m}/0.3}$ & --- & $0.87\pm0.18$ & $0.72\pm0.08$ \\
$\widehat{E_{\rm G}}$ & --- & --- & $0.33\pm0.10$ \\
$f\sigma_8(z=0.57)$ & $0.53\pm0.14$ & $0.50\pm0.86$ & $0.51\pm0.12$ \\
$\sigma_8(z=0)$ & $0.78\pm0.26$ & $0.70\pm0.12$ & $0.73\pm0.08$ \\
\hline

\end{tabular}
\caption{Best-fit and derived parameters obtained by fitting the RSD only, GGL only, and their combination.}
\label{tab:fitresults}
\end{table*}

\subsection{Comparison with other measurements}

\ch{From MCMC, we can derive new parameter constraints, defined as combination of single parameters.}  
In particular, we look at the quantity $S_8 = \sigma_8 \sqrt{\Omega_{\rm m}/0.3}$, very common in gravitational lensing analyses. We find $S_8 = 0.72\pm0.08$, which is in agreement with the value estimated in L17, but $2-3\sigma$ smaller than the CMB measurements $S_8 = 0.832\pm0.013$ \citep{planck2018}. Similarly, our estimate of $\sigma_8 = 0.73\pm0.08$ is $2-3\sigma$ smaller than the measurement $\sigma_8=0.8111\pm0.0060$ from the Planck collaboration 2018. Our results are also in agreement with KiDS shear peaks statistics $S_8 = 0.75\pm 0.059$ \citep{martinet2018,shan2018}, KIDS tomographic weak lensing $S_8 = 0.745\pm0.039$ \citep{hildebrandt2017}, and DES cosmological constraints from weak-lensing and clustering $S_8 = 0.783^{+0.021}_{-0.025}$. We note that our fit only performed with RSD measurements yield an estimate of $\sigma_8 = 0.78\pm0.26$, in better agreement with Planck estimates.

The linear galaxy bias parameter is known to be degenerate with the cosmological parameters $\Omega_{\rm m}$ and $\sigma_8$. In our fitting procedure, we assume a uniform prior on $b_1$ between 1 and 3, which largely encompasses the expected value for the CMASS sample. In their clustering analyses, \cite{gilmarin2017} found $b_1 \sigma_8 (z=0.57) = 1.237\pm 0.011$, and \cite{chuang2013} found $b_1 \sigma_8 (z=0.57) = 1.18\pm 0.14$. We find $b_1 \sigma_8 (z=0.57) = 1.256\pm0.097$ in full agreement with these previous measurements. Marginalizing over $\sigma_8$, we find $b_1 = 2.33\pm0.33$, in agreement with \cite{white2011} and subsequent analyses \citep[e.g.][]{ho2012,nuza2012,rodriguez2016}.

Our model also contains 2$^{nd}$ order biasing term, but our estimated value $b_2 = -0.04\pm0.53$ is not sufficient to discuss the non-linearity of the CMASS sample. Note that \cite{gilmarin2017} found $b_2 = 0.606\pm0.069$, which is in agreement with us. 

Finally, we also include Alcock-Paczynski effect in our model, but found no significant constrain given the data, $\alpha = 1.01\pm0.05$ and $\epsilon = 0.00\pm0.05$. Note that no significant constrain could either be obtained by \cite{gilmarin2017} with the full CMASS DR12 sample. 

To conclude, we demonstrated the effectiveness of combining RSD and GGL to break the degeneracies between the amplitude of the large-scale structure fluctuations $\sigma_8$ and their growth rate $f$ at redshift $z=0.57$. We also found that the constraints on the cosmic matter density $\Omega_{\rm m}$, usually derived with weak-lensing, could be significantly improved by combining with RSD. Given the data, our measurements are still in agreement with Planck predictions. 

\subsection{Measuring $E_{\rm G}$}
\label{sec:eg}

To corroborate the information obtained with the analysis in the previous section and probe any deviation to $\Lambda$CDM-GR predictions, we estimate $E_{\rm G}$, as defined in \cite{reyes2010}. The $E_{\rm G}$ estimator is function of projected scale $r_{\rm p}$, and is defined as \citep{zhang2007}

\begin{equation}
E_{\rm G}(r_{\rm p}) = \frac{1}{\beta} \frac{\Upsilon_{{\rm gm}} (r_{\rm p})}{\Upsilon_{{\rm gg}} (r_{\rm p})}\,.
\label{eq:estimator}
\end{equation}

\noindent This estimator is particularly interesting because it apparently just relies on observations. However, we show in the following that this might not be the case in practice. 

Indeed, the $E_{\rm G}$ estimator suffers from a few downsides. First, it relies on a previous determination of $\beta$. \ch{However, statistical and systematic} error propagation into $E_{\rm G}$ error might be awkward, unless proper \ch{correction terms} and covariance matrices are determined from ad-hoc mock catalogues of lensing and clustering. \ch{Although seldom the case in the past, this becomes more and more common \citep{blake2016,amon2017,singh2018}.} \st{We do this in the present work, but this is seldom the case. Also, it usually happens that the correlation scales probed to determine $\beta$ are different from the scales at which $E_{\rm G}$ is estimated. }\cite{singh2018} and \cite{amon2017} \st{have assumed their $\beta$ estimates are scale independent.}

Second, it is assumed that galaxy bias is linear, scale-independent and the galaxy density field is fully correlated to the underlying matter density field, i.e. the cross-correlation factor $r_{cc} = 1$. \st{This point is closely related to the previous point about $\beta$, since $\beta = f /b$.}  Of course, these assumptions hold in the linear regime, but the scale at which they break depends on the galaxy sample. Using CMASS mock catalogues, several authors have shown that they hold in the range $5 < r_{\rm p} < 60 h^{-1}\ {\rm Mpc}$ \cite{baldauf2010,white2011,amon2017,singh2018}. This depends on the requested precision on the model though, and recent works have proposed to take non-linearity and other effects into account with normalizing functions derived from simulations \citep{alam2016, singh2018}. \ch{The multiplication of these correction terms nonetheless tend to reveal the limitation of the $E_{\rm G}$ estimator.} 

\ch{\cite{martapinho2018} note that $E_{\rm G}$ depends not only on gravity but also on the background (e.g. quantified with the matter density $\Omega_{\rm m0}$ in $\Lambda$CDM). Although it is always possible to predict $E_{\rm G}$ for different cosmological models \citep[see e.g.][in which predictions are computed for $\Lambda$CDM, Flat DGP, f(R) gravity, TeVeS/MOND]{zhang2007}, a discrepancy with the observations therefore does not specifically point to a failure of General Relativity, but can also be attributed to the background. In this respect, they claim that an estimator such as $\eta$, based on independent estimates of $f\sigma_8(z)$, $H(z)$, $E_{\rm G}$ might be more appropriate. To our point of view, adjusting an actual model including modified gravity parameters might be as helpful. }

In spite of these limitations, $E_{\rm G}$ has become quite popular recently, mostly because of the advent of wide field imaging and spectroscopic surveys. It has been measured several times, but no significant deviation from $\Lambda$CDM-GR has been found so far.  In particular with the CMASS sample at redshift z$=0.57$, \cite{amon2017} found $E_{\rm G} = 0.26\pm0.08$, \cite{blake2016} found $E_{\rm G} = 0.30\pm0.07$, \cite{pullen2016} found $E_{\rm G} = 0.24\pm0.06$, \cite{alam2016} found $E_{\rm G} = 0.42\pm0.06$, and \cite{singh2018} found $E_{\rm G}=0.39\pm0.05$. \ch{The dispersion in the estimates reveal that the method is probably not fully mature yet, and deserves further investigation, in particular regarding the observational biases, such as photometric redshifts.}

\begin{figure}
\includegraphics[width=\linewidth]{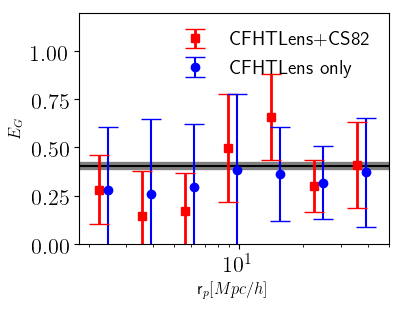}
\caption{Measurement of $E_{\rm G}$ with combined constraints in the fields CFHT-Stripe 82  and CFHTLs W1, W3 and W4. The horizontal black line indicates the Planck TT,TE,EE+lowE 2018 prediction. Note that CFHT-S82 data help shrink the error bars by about 30\%. CFHTLens points have been shifted rightwards for clarity.}
\label{fig:eg}
\end{figure}

Figure~\ref{fig:eg} shows our measurements of $E_{\rm G}$ as a function of scale. We estimate $\beta = 0.41\pm0.15$ from our fit to the RSD measurements only. Although difficult to compare because people use different models, this value is \st{1$\sigma$ larger than} \ch{larger but statistically consistent with} the one found with the full CMASS sample $\beta = 0.34\pm0.02$ \ch{from \cite{amon2017}.
For each bin of $E_{\rm G}(r_p)$, we add in quadrature the errors on the ratio $U^i = \Upsilon_{\rm gm}^i / \Upsilon_{\rm gg}^i$ derived from the data and the error on $\beta = f/b_1$ derived from the fit, with the following chain rule formula

\begin{equation}
\frac{\hat{C}(E_{\rm G}^i)}{E_{\rm G}^i E_{\rm G}^i} = \frac{\hat{C}(U^i)}{U^i U^i} +  \frac{\hat{C}(\beta)}{\beta^2} + 2\sqrt{\frac{\hat{C}(\beta)}{\beta^2}} \sqrt{\frac{\hat{C}(U^i)}{U^i U^i}} \rho(\beta,U^i)
\end{equation}

\noindent Using the MCMC samples from the fit of the GGL and RSD measurements, we use our model to reconstruct the ratio $U^i$. We also determine the correlation coefficients $\rho(\beta,U^i) \sim 0.3$, i.e. $\beta$ and $U^i$ are significantly correlated.  } 

We average in the scale range $10 < r_{\rm p} < 60 h^{-1}\ {\rm Mpc}$, and find $E_{\rm G} = 0.48\pm0.15$ for CFHTLens field only, and $E_{\rm G} = 0.43\pm0.11$ for CFHTLens and CS82 fields combined, i.e. a 30\% improvement in precision for a 100\% increase in area. In the average $E_{\rm G}$ calculation, we consider the full covariance matrix between the $E_{\rm G}$ points estimated from our simulations in section~\ref{sec:simu}.  Note finally, that our current precision does not justify applying scale-dependent bias, redshift weighting or integration window corrections, since their effect is less than 5\% at the scales we consider \cite[see][]{alam2016,singh2018}. 

To put our measurement in context, we collected the $E_{\rm G}$ measurements at different redshifts from the literature in Figure~\ref{fig:egz}. Overall, we observe a trend of $E_{\rm G}$ values lower than predicted by Planck 2018. In the Appendix, we forward model the $E_{\rm G}$ signal based on the MCMC samples output from the joint fit of the GGL and RSD measurements on mocks. Figure~\ref{fig:eg_r10} shows that the probability distribution function of the $E_{\rm G}$ estimator is skewed towards low values. Taking its mean value then necessarily leads to biased-low estimation of $E_{\rm G}(r_{\rm p})$. This result confirms the previous claim from \cite{alam2016}, and might also explain why so many $E_{\rm G}$ measurements are below the Planck 2018 predictions. 

\begin{figure*}
\includegraphics[width=\linewidth]{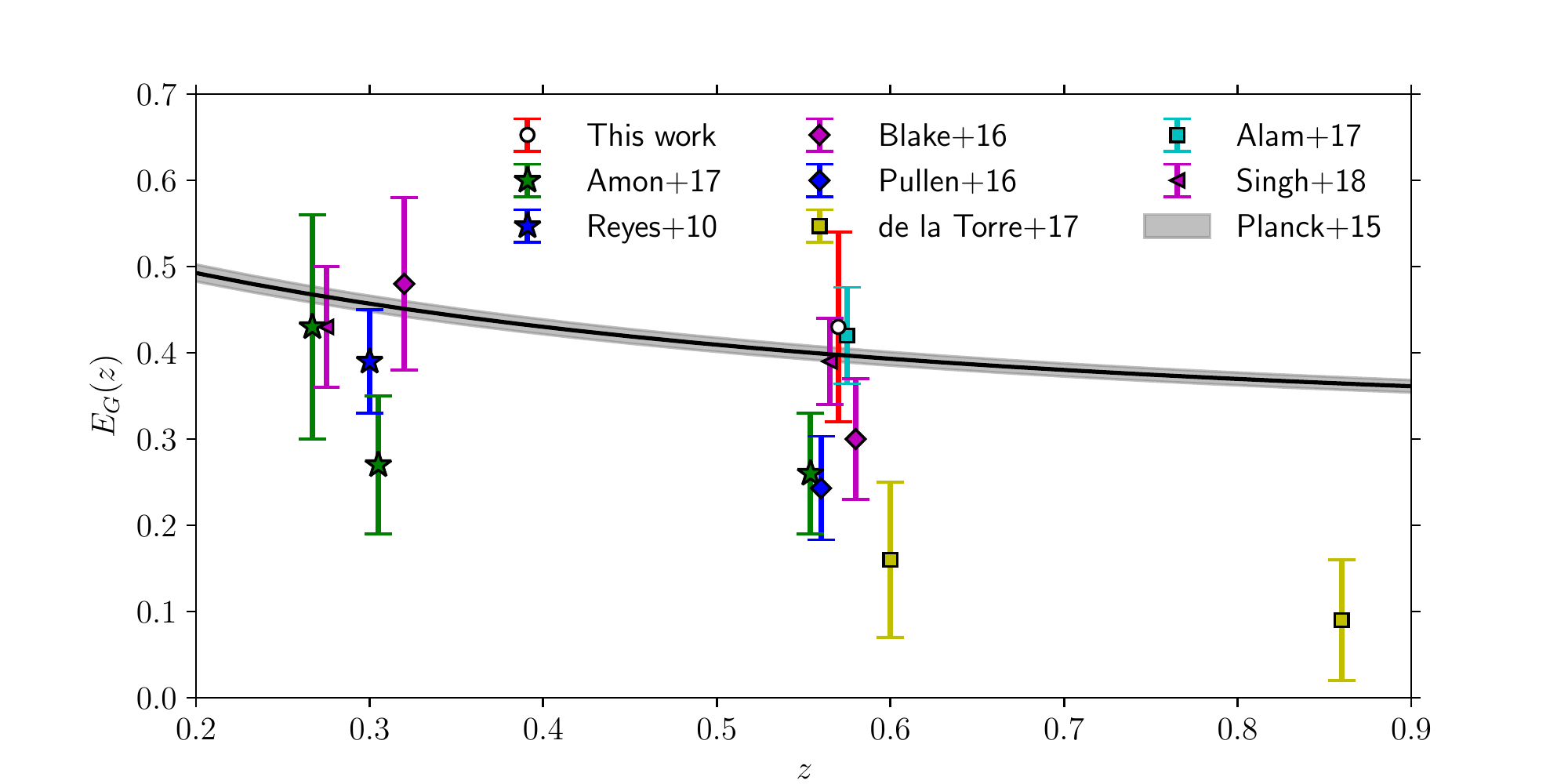}
\caption{Measurements of $E_{\rm G}$ at different redshifts. Effective redshifts of the measurements have been slightly shifted for clarity.}
\label{fig:egz}
\end{figure*}

\section{Conclusion}

Understanding the current acceleration of the Universe's expansion is one of the major goal of cosmology today. The combination of GGL and RSD is a privileged avenue to distinguish the effect of gravity due to large-scale structures, and the effect of some scalar field on the background expansion rate. 

In this work, we have demonstrated the power of this combination applied to the well studied CMASS galaxy sample at the effective redshift $z=0.57$. Using a comprehensive set of lensing and galaxy mock catalogues, we have investigated several sources of systematic biases, and determined the confidence limits for our datasets. In particular, we have found that thanks to spectroscopic data, we could correct the bias due to photometric redshift uncertainty for galaxies brighter than $i_{\rm AB} < 22.5$, and $i_{\rm AB} < 24$ in our CFHT-S82 and CFHTLens weak lensing catalogues respectively. 
These conservative magnitude cuts allow us to match our GGL measurements in the CFHT-S82 and CFHTLS fields, and with predictions based on model fitting of RSD measurements. Nonetheless, they highlight the crucial need of spectroscopic redshifts for faint galaxies.

Building on this encouraging result, we pursue a cosmological analysis of the combined dataset. Thanks to the joint GGL and RSD constraints, we efficiently break the degeneracy between galaxy bias $b_1$, \ch{matter density $\Omega_{\rm m}$, matter power spectrum amplitude $\sigma_8$} and the structure growth rate $f$ at $z=0.57$. We find astrophysical CMASS parameters and cosmological parameters in agreement with measurements previously obtained by other authors \citep{white2011,beutler2014,chuang2013,gilmarin2017,joudaki2017}, and with Planck 2018 predictions in the frame of the $\Lambda$CDM-GR model.

Finally, we combine GGL and RSD measurements to estimate $E_{\rm G}$. By averaging in the range of scales $10 < r_{\rm p} < 60 h^{-1}\ {\rm Mpc}$, we find $E_{\rm G}(z=0.57) = 0.43\pm0.11$, which is in perfect agreement with Planck 2018 prediction $E_{\rm G} = 0.40$. Also, we use our mocks to characterize the statistical properties of $E_{\rm G}$, and find that it has an asymmetric probability distribution, which tends to underestimate its mean value. This might explain part of the low values found in previous analysis. \st{Using end-to-end simulations,} We \ch{also} find that \st{a proper combined fit of GGL and RSD measurements,} \ch{the reconstructed value of $E_{\rm G} = \Omega_{\rm m0} /f$ derived from the fit of the GGL and RSD measurements results in a value with smaller errors bars than the one obtained directly from the data. More importantly, it naturally includes the cross-correlation terms between $\beta$ and $\Upsilon_{gg}$.} 

Back in 2012, \citeauthor{gaztanaga2012} was already advocating that overlapping lensing and spectroscopic surveys were 100 times more constraining on dark energy equation of state, and cosmic growth history parameter $\gamma$. Although it might not be the cleanest way to test gravity, the recent progress on estimating  $E_{\rm G}$ at different redshifts with different tracers comes as a confirmation. 
In the future, wider imaging and spectroscopic surveys will result in very tight constraints on cosmological parameters. In contrast, it will probably take us more time to take full profit of smaller but deeper imaging surveys. Deep imaging surveys are helpful for many reasons, but also introduce additional systematic errors on the lensing side, in particular with respect to blending \citep{harnois2018,euclidmartinet2019}. Nonetheless, both strategies lead to very exiting perspectives regarding our understanding of the dark sector.

\section{Acknowledgements}
Based on observations obtained with MegaPrime/MegaCam, a joint project of CFHT and CEA/DAPNIA, at the Canada-France-Hawaii Telescope (CFHT), which is operated by the National Research Council (NRC) of Canada, the Institut National des Science de l'Univers of the Centre National de la Recherche Scientifique (CNRS) of France, and the University of Hawaii.  The Brazilian partnership on CFHT is managed by the Laborat\'orio Nacional de Astrof\'isica (LNA). We thank the support of the Laborat\'orio Interinstitucional de e-Astronomia (LIneA). We thank the CFHTLenS team. Funding for SDSS-III has been provided by the Alfred P. Sloan Foundation, the Participating Institutions, the National Science Foundation, and the U.S. Department of Energy Office of Science. The SDSS-III web site is \url{http://www.sdss3.org/}.
SDSS-III is managed by the Astrophysical Research Consortium for the Participating Institutions of the SDSS-III Collaboration including the University of Arizona, the Brazilian Participation Group, Brookhaven National Laboratory, Carnegie Mellon University, University of Florida, the French Participation Group, the German Participation Group, Harvard University, the Instituto de Astrofisica de Canarias, the Michigan State/Notre Dame/JINA Participation Group, Johns Hopkins University, Lawrence Berkeley National Laboratory, Max Planck Institute for Astrophysics, Max Planck Institute for Extraterrestrial Physics, New Mexico State University, New York University, Ohio State University, Pennsylvania State University, University of Portsmouth, Princeton University, the Spanish Participation Group, University of Tokyo, University of Utah, Vanderbilt University, University of Virginia, University of Washington, and Yale University. The BigMDPL simulation has been performed on the SuperMUC supercomputer at the Leibniz-Rechenzentrum (LRZ) in Munich, using the computing resources awarded to the PRACE project number 2012060963. We thank the Red Espa\~nola de Supercomputaci\'on for granting us computing time in the Marenostrum Supercomputer at the BSC-CNS where part of the analyses presented in this paper have been performed. 
We thank the support of the OCEVU Labex (Grant N$^{\rm o}$ ANR-11-LABX-0060) and the A*MIDEX project (Grant N$^{\rm o}$ ANR-11-IDEX-0001-02) funded by the Investissements d'Avenir French government program managed by the ANR. We also acknowledge support from the ANR eBOSS project (ANR-16-CE31-0021) of the French National Research Agency. 
CG acknowledges support from Centre National d'Etudes Spatiales, Italian  Ministry of  Foreign Affairs  and International  Cooperation Directorate General  for Country  Promotion (Project ``Crack the lens?''),  from the agreement ASI n.I/023/12/0 ``Attivit\`a relative  alla  fase  B2/C  per   la  missione  Euclid'' and from the Italian Ministry for Education, University  and  Research  (MIUR)  through the  SIR  individual  grant SIMCODE (project number RBSI14P4IH). JPK acknowledges support from the the ``Cosmology with 3D-Maps of the Universe'' SNF grant \#175751. GY acknowledges financial support from MINECO/FEDER under project grant  AYA2015-63810-P

\vspace{1cm}


\appendix
\section{Weak lensing systematics tests}

\paragraph{masking} In order to assess the impact of missing tiles and large scale masking (e.g. due to very bright stars), we compute the density of CS82 galaxies on a grid with pixel size $\sim 1\ {\rm deg}$. Then, we randomly draw mock galaxies in the field such that the overall redshift distribution and total number of sources matches observations. Finally, we down-sample this catalogue according the density fluctuations attributed to masking. We find that masking increases the statistical noise in the GGL measurement by about 20\% at all scales. However we could not identify any obvious systematic bias related to masking.

\paragraph{photometric redshifts bias} 
\label{sec:zphbias} \cite{mandelbaum2008} and \cite{nakajima2012} proposed an alternative method to estimate the bias introduced by photometric redshifts on galaxy-galaxy lensing measurements. They proposed to estimate the bias $b_z(z_{lens})$ between photometric redshifts $\Delta \tilde{\Sigma}$ and spectroscopic redshifts $\Delta \Sigma$ measurements, 

\begin{equation}
1+b_z(z_{\rm lens}) = \frac{\Delta \tilde{\Sigma}}{\Delta \Sigma} = \frac{\sum_j w_j \Sigma_{\rm cr}^{-1} \tilde{\Sigma}_{\rm cr}^{-1}}{\sum_j w_j  \tilde{\Sigma}_{\rm cr}^{-2}}\,.
\end{equation}

\noindent The summation is performed over the subset of source galaxies with both spectroscopic and photometric redshifts. We adapted the original expression from \cite{mandelbaum2008} such that the inverse critical densities $\Sigma_{\rm cr}^{-1} = \frac{4\pi G}{c^2} D_L (1 - \frac{D_L}{D_S})$ converges to zero when the source redshift becomes smaller than $z_{\rm lens}$. $w_j$ is the weight on source galaxy $j$ in the lensing catalogue. In order to estimate the effective bias on our galaxy-galaxy lensing measurements with CMASS galaxies, we need to integrate the bias function $b_z(z_{\rm lens})$ over the CMASS redshift distribution $p(z_{\rm lens})$ such that 

\begin{equation}
\langle b_z \rangle = \frac{\int \d z_{\rm lens}\ p(z_{\rm lens}) \tilde{w}_l(z_{\rm lens})  b_z(z_{\rm lens})}{\int \d z_{\rm lens}\ p(z_{\rm lens}) \tilde{w}_l(z_{\rm lens})}
\end{equation}

\noindent where the weight on each lens place $\tilde{w}_l = D_L^{-2} (1+z_{\rm lens})^{-2} \sum_j w_j \Sigma_{\rm cr}^{-2}$ is correcting for the fact that the number of sources involved in a given aperture in physical coordinates includes more objects at lower than at higher redshifts. We bootstrapped our catalogs to estimate the uncertainties on our bias estimates.

For this measurement, we used VVDS ($i_{\rm AB} < 22.5$, \cite{garilli2008}), DEEP2 ($R_{\rm AB} < 24.1$, \cite{newman2013}), PRIMUS ($i_{\rm AB} < 23.5$, \cite{coil2011}), VIPERS ($i_{\rm AB} < 22.5$, \cite{guzzo2014}) and SDSS-DR13 spectroscopic redshifts, that we matched to our lensing sources in our 4 fields. On Stripe 82, we obtained $b_z = -0.028\pm0.006$, $b_z=-0.131\pm0.004$ and $b_z=-0.082\pm0.004$ for BPZ, Neural Network or LePhare codes respectively. With CFHTLens, we obtained $b_z=+0.003\pm0.003$, $b_z = -0.014\pm0.004$ and $b_z=+0.022\pm0.003$ on fields W1, W3 and W4 respectively. 

We also adapted our code to assess the improvement obtained by using the photometric redshift probability of each source galaxy $p(z_s)$ instead of maximum likelihood values. The critical densities then become

\begin{equation}
\tilde{\Sigma}_{\rm cr}^{-1} =  \frac{4\pi G}{c^2} D_L \int \d z_s\ p(z_s) (1 - \frac{D_L}{D_S})\,.
\end{equation}

On field Stripe 82, we found that using the full probability $p(z_s)$ halved the bias obtained with LePhare code to $b_z = -0.031\pm0.005$. Nonetheless, using the best-fit redshifts provided by BPZ still yields the smallest bias.

\paragraph{catastrophic photometric redshifts} In order to assess the impact of catastrophic redshifts on the lensing measurements, we computed the 2 dimensional probability $p(z_{\rm LP}|z_{\rm ANNZ})$ of obtaining a photometric redshift with Le Phare given a photometric redshift obtained with Neural Network. Assuming this later to be the true redshift, we degraded the true redshifts in our mocks in order to reproduce the catastrophic outlier effects. Overall, we found that catastrophic redshifts could bias the lensing signal by about $b_z=+0.03$. This is in agreement with our estimations above with spectroscopic redshifts, as well as the estimates found in  \cite{leauthaud2017}.

\begin{figure}
\centering
\includegraphics[width=\linewidth]{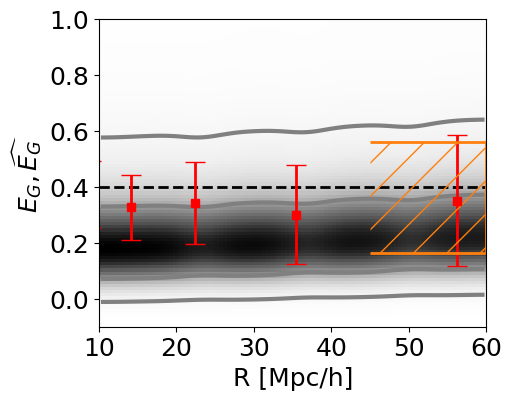} 
\caption{Recovered signal in the mocks, when $E_{\rm G} = \frac{\Upsilon_{\rm gm}}{\beta \Upsilon_{\rm gg}}$  (grey shared area with $1\sigma$ and $2\sigma$ C.L. contours) and $\widehat{E_{\rm G}} = \Omega_{\rm m} /f$ (orange $1\sigma$ shaded area) are computed from the MCMC samples, and when $E_{\rm G} = \frac{\Upsilon_{\rm gm}}{\beta \Upsilon_{\rm gg}}$ is directly estimated from the mocks (red data points with $1\sigma$ error bars). In this latter case, we take $\beta = 0.84/2.13 = 0.39$, as obtained from a previous fit of our model to the RSD-only measurements in the mocks. Both definitions are in agreement with the value of $\Omega_{\rm m} /f$ computed using the Planck cosmology 2018 of the simulation (black dashed line). Measurements are 
performed with $R_0 = 1.0\ h^{-1} {\rm Mpc}$, $s_{\rm min} = 17.8\ h^{-1} {\rm Mpc}$ and no tapering.}
\label{fig:eg_r10}
\end{figure}

\paragraph{asymmetric posterior on $E_{\rm G}$}  It is very typical that observational estimators obtained from a ratio of observables have asymmetric probability distribution function. Indeed using our simulations, we found that $E_{\rm G}$ is systematically lower than $\Lambda$CDM-GR predictions, with a long tail towards larger values of $E_{\rm G}$, as shown in Figure~\ref{fig:eg_r10}. When applying the usual $E_{\rm G} =\frac{\Upsilon_{\rm gm}}{\beta \Upsilon_{\rm gg}}$ estimator on our mocks, we also find mean values smaller than expected, although still in statistical agreement. Note finally that 1$\sigma$ constraints are tighter with the fit than with the usual $E_{\rm G}$ estimator. Given that \cite{amendola2013} have demonstrated that $E_{\rm G}$ is probably not as effective as anticipated in some particular cosmological models, and \cite{alam2016} and \cite{singh2018} have started to apply bias corrections to this estimator based on mock catalogues built with fiducial cosmological models, we think it might be as efficient and clean to fit the correlation functions\ch{, and derive $E_{\rm G}$ by marginalizing over the model parameters, or simply compare $\Omega_{\rm m0}$ and $f$ to their $\Lambda$CDM+GR predictions. }

\begin{figure*}
\includegraphics[width=\linewidth]{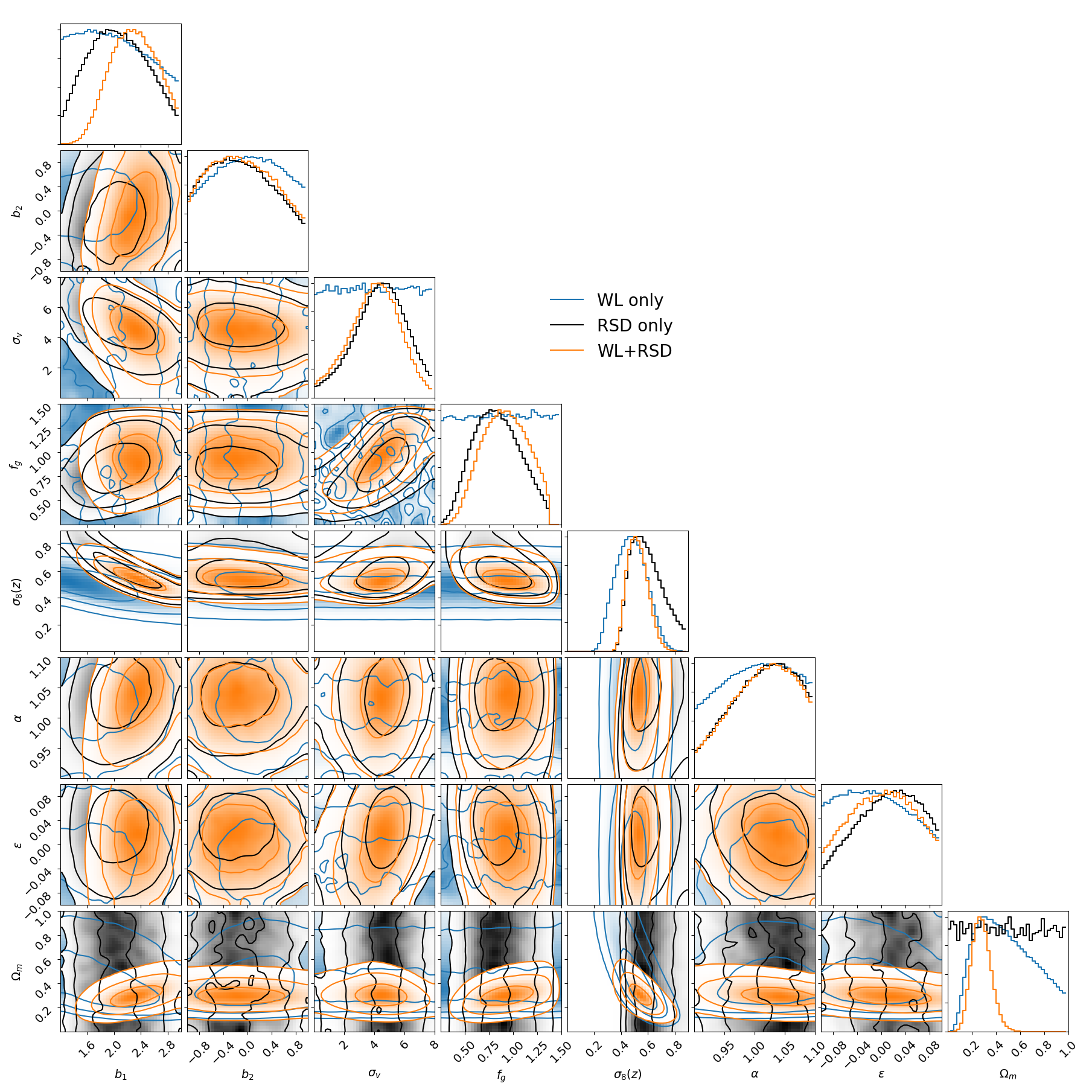}
\caption{Contours at 1, 2 and 3$\sigma$ C.L. of all our model parameters estimated with the RSD only, GGL only and their combination. In all cases, we set $s_{\rm min} = 17.8h^{-1}\ {\rm Mpc}$ and $R_0 = 1.0 h^{-1}\ {\rm Mpc}$ (see text for details).}
\label{fig:corner}
\end{figure*}

\end{document}